\documentclass[reprint,aps,
        nolongbibliography,
		citeautoscript
		]{revtex4-2}

\usepackage{amsmath}
\usepackage{booktabs}
\usepackage{color}
\usepackage{graphicx}
\usepackage[italic]{hepnames}
\usepackage{hyperref} % necessary?
\usepackage{lineno}
\usepackage{physics}
\usepackage[separate-uncertainty=true]{siunitx}
\usepackage{url} % necessary?
\usepackage{xfp}
\usepackage{orcidlink}

\def\verbatim@font{\normalfont\ttfamily}

\newif\ifnaturalunits
\naturalunitstrue
\newcommand{\naturalunits}[1]{\ifnaturalunits\else #1\fi}

\newcommand{\tauBzResult}{\SIStatSyst{1.499}{0.013}{0.008}{ps}}
\newcommand{\dmdResult}{\SIStatSyst{0.516}{0.008}{0.005}{ps^{-1}\naturalunits{\hbar/c^2}}}

\newcommand{\sub}[1]{\ensuremath{_{\text{#1}}}\xspace}
\newcommand{\SIStatSyst}[4]{\ensuremath{(#1 \pm #2 \pm #3)\,\si{#4}}\xspace}

\newcommand{\belleii}{\ensuremath{\text{Belle~II}}\xspace}

\newcommand{\tauBz}[1][]{\ensuremath{\tau^{#1}_{\hspace{-0.22em}\PBzero}}\xspace}
\newcommand{\dmd}{\ensuremath{\Delta m_{\hspace{-0.11em}\Pdown}}\xspace}

\newcommand{\dectime}{\ensuremath{\Delta t}\xspace}
\newcommand{\measdectime}{\ensuremath{\dectime_{\ell}}\xspace}
\newcommand{\dectimeunc}{\ensuremath{\sigma_{\!\dectime_{\ell}}}\xspace}

\newcommand{\declength}{\ensuremath{\ell}\xspace}
\newcommand{\avgdectime}{\ensuremath{\bar{t}}\xspace}
\newcommand{\dectimeoffset}{\ensuremath{\delta t}\xspace}

\newcommand{\cosdmd}{\ensuremath{\cosine(\dmd\dectime\naturalunits{\tfrac{\hbar}{c^2}})}\xspace}

\newcommand{\resfunc}{\ensuremath{R}\xspace}
\newcommand{\gausfunc}{\ensuremath{G}\xspace}
\newcommand{\tailsub}{\ensuremath{t}}
\newcommand{\OLsub}{\ensuremath{\text{OL}}}

\renewcommand{\APquark}{\ensuremath{\bar\Pquark}\xspace}
\renewcommand{\APB}{\ensuremath{\bar{\PB}}\xspace}
\renewcommand{\PBzero}{\ensuremath{\PB^{0}}\xspace}
\renewcommand{\APBzero}{\ensuremath{\bar{B}{}^0}\xspace}
\renewcommand{\APDzero}{\ensuremath{\bar{D}{}^0}\xspace}
\renewcommand{\Ppizero}{\ensuremath{\Ppi^{0}}\xspace}
\renewcommand{\PUpsilonFourS}{\ensuremath{\Upsilon(4\text{S})}\xspace}

\newcommand{\PBsig}{\ensuremath{\PB_{\text{sig}}}\xspace}
\newcommand{\PBtag}{\ensuremath{\PB_{\text{tag}}}\xspace}

\newcommand{\PDstarminus}{\ensuremath{\PD^{*-}}\xspace}
\newcommand{\PDorDstarminus}{\ensuremath{\PD^{(*)-}}\xspace}

\begin{document}

%\setfootnotestyle{super}

%%%%%%%%%%%%%%%%%%%%%%%%%%%%%%%%%%%%%%%%%%%%%%%%%%
\title{Measurement of the \PBzero lifetime and flavor-oscillation
  frequency using hadronic decays reconstructed in  2019-2021 \belleii data}

\noaffiliation

%%%%%%%%%%%%%%%%%%%%%%%%%
\vspace*{-3\baselineskip}
\resizebox{!}{3cm}{\includegraphics{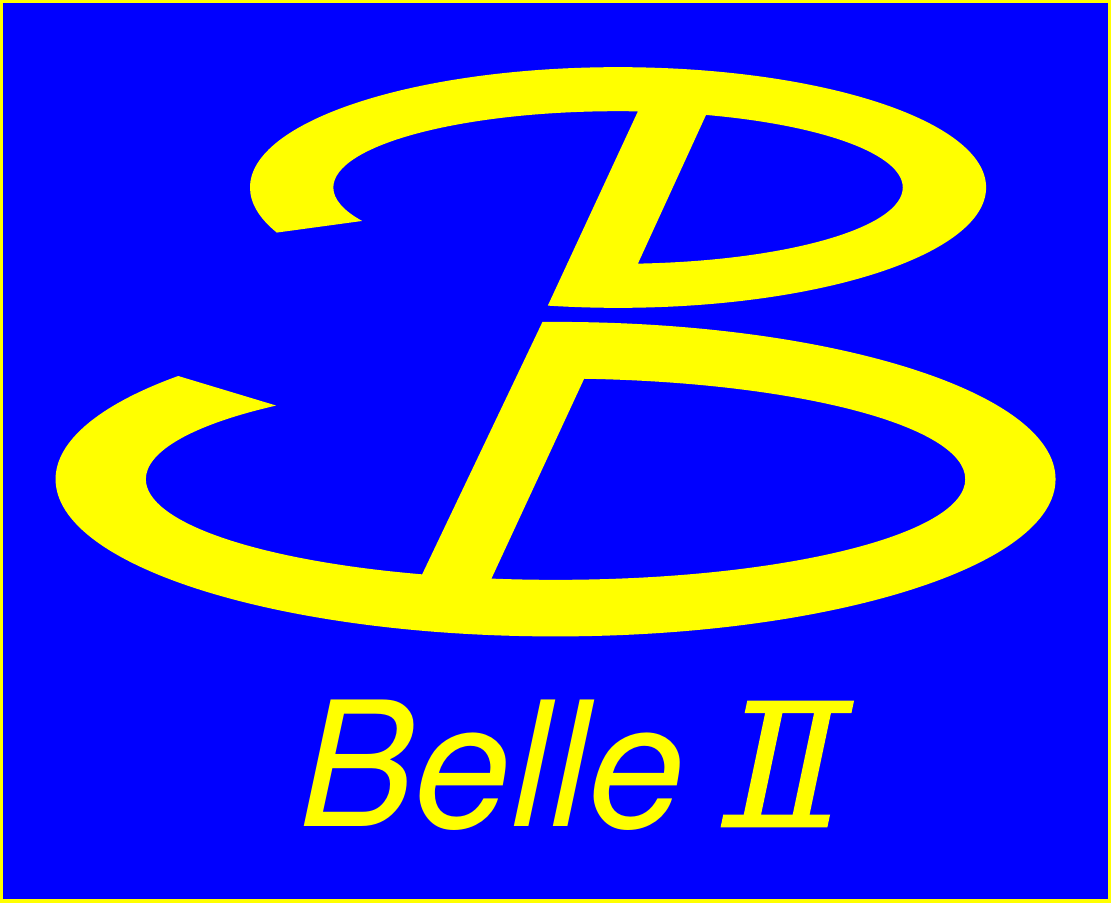}}
 
\begin{flushright}
  February 24, 2023
\end{flushright}

%\input{authors-plain.tex}
%%% Paper:    B0 lifetime and flavor-oscillation frequency
%%% Journal:  Physical Review D Letters
%%% Contacts: Thibaud Humair, Justin  Skorupa, Jacob Kandra, Radek Zlebcik,
%%%           Caspar Schmitt, Vladimir Shekelyan
%%% ====================================================================
%%% Use \input{pub016-orcid} to insert this material into your latex file.
  \author{F.~Abudin{\'e}n\,\orcidlink{0000-0002-6737-3528}} % 2250
  \author{I.~Adachi\,\orcidlink{0000-0003-2287-0173}} % 2590
% \author{K.~Adamczyk\,\orcidlink{0000-0001-6208-0876}} % 2239
  \author{L.~Aggarwal\,\orcidlink{0000-0002-0909-7537}} % 10083
% \author{P.~Ahlburg\,\orcidlink{0000-0002-9832-7604}} % 2367
  \author{H.~Ahmed\,\orcidlink{0000-0003-3976-7498}} % 11323
% \author{J.~K.~Ahn\,\orcidlink{0000-0002-5795-2243}} % 7423
  \author{H.~Aihara\,\orcidlink{0000-0002-1907-5964}} % 2223
  \author{N.~Akopov\,\orcidlink{0000-0002-4425-2096}} % 9443
  \author{A.~Aloisio\,\orcidlink{0000-0002-3883-6693}} % 2194
% \author{F.~Ameli\,\orcidlink{0000-0001-5435-0450}} % 4683
% \author{L.~Andricek\,\orcidlink{0000-0003-1755-4475}} % 2098
  \author{N.~Anh~Ky\,\orcidlink{0000-0003-0471-197X}} % 2218
  \author{D.~M.~Asner\,\orcidlink{0000-0002-1586-5790}} % 4684
  \author{H.~Atmacan\,\orcidlink{0000-0003-2435-501X}} % 2538
% \author{V.~Aulchenko\,\orcidlink{0000-0002-5394-4406}} % 8183
% ...
% 13403
\collaboration{The Belle II Collaboration}

%%%%%%%%%%%%%%%%%%%%%%%%%

%%%%%%%%%%%%%%%%%%%%%%%%%%%%%%%%%%%%%%%%%%%%%%%%%%
\begin{abstract}

    We measure the \PBzero lifetime and flavor-oscillation frequency 
    using \HepProcess{\PBzero\to\PD^{(*)-}\Ppiplus} decays 
    collected by the Belle II experiment in asymmetric-energy $\APelectron\Pelectron$
     collisions produced by the SuperKEKB collider operating at the \PUpsilonFourS resonance.
    We fit the decay-time distribution of signal decays, where the initial
	  flavor is determined by identifying the flavor of the other $\PB$
    meson in the event. 
    The results, based on $33000$ signal decays reconstructed in a data sample
    corresponding to \SI{190}{fb^{-1}},
    are
  \begin{alignat*}{1}
    \tauBz &= \tauBzResult\\
    \dmd   &= \dmdResult,
  \end{alignat*}
  where the first uncertainties are statistical and the second
  are systematic. 
  These results are consistent with the world-average values.
  
\end{abstract}
%%%%%%%%%%%%%%%%%%%%%%%%%%%%%%%%%%%%%%%%%%%%%%%%%%


%%%%%%%%%%%%%%%%%%%%%%%%%
\maketitle

Knowledge of the \PBzero lifetime \tauBz, and the flavor-oscillation frequency \dmd, 
allows us to test both the QCD theory of strong interactions at low energy 
and the Cabibbo-Kobayashi-Maskawa (CKM) theory of weak interactions~\cite{Lenz:2014jha, Dowdall:2019bea}.
The Belle, Babar, 
and LHCb collaborations have measured 
\tauBz and \dmd to comparable 
precision~\cite{Belle:2004hwe,BaBar:2005laz,LHCb:2014qsd,LHCb:2016gsk}.
Additionally, the CMS, ATLAS, D0 and CDF collaborations have measured 
\tauBz to similar precision~\cite{CMS:2017ygm,ATLAS:2012cvl,D0:2014ycx,CDF:2010ibe}.
LHCb's  measurements, 
$\tauBz = \SIStatSyst{1.524}{0.006}{0.004}{ps}$ and
$\dmd= \SIStatSyst{0.5050}{0.0021}{0.0010}{ps^{-1}\naturalunits{\hbar/c^2}}$\ifnaturalunits\else\fi,
are the most precise to-date~\cite{LHCb:2014qsd,LHCb:2016gsk}.\footnote{We use a system 
of units in which
  $\hbar=c=1$ and mass and frequency have the same
  dimension.}
When two uncertainties are given, the first is statistical and the
second is systematic. 

Here we report a new measurement of \tauBz and \dmd using 
hadronic decays of \PBzero mesons reconstructed in a \SI{190}{fb^{-1}}  data set collected
by the Belle II experiment at the SuperKEKB asymmetric-energy
$\APelectron\Pelectron$ collider. The data were collected between 2019 and 2021. The \PBzero mesons are produced in 
the
$\APelectron\Pelectron\to\PUpsilonFourS\to\PB\APB$ process, 
where $\PB$ indicates a $\PBzero$ or a $\PBplus$. 
Our data set
contains approximately 200~million such events.
Our measurement tests the ability of Belle~II to precisely measure \PBzero meson 
decay times and also identify the initial flavor of the decaying \PBzero; 
such capabilities are crucial for measuring decay-time-dependent \textit{CP} violation
 and determining $\phi_1$ and $\phi_2$, 
 two of the three angles of the \PBzero 
 CKM unitarity triangle.\footnote{Another naming convention, with 
 $\beta  \equiv \phi_1$ and $\alpha\equiv\phi_2$, is also used in the literature.}
 Examples of measurements of  $\phi_1$ and $\phi_2$ 
 are found in Refs.~\cite{Belle:2012paq,Belle:2015xfb}.

The flavor of a neutral \PBzero or \APBzero meson oscillates  with frequency
$\dmd\naturalunits{\hbar/c^2}$ before it decays. The probability 
density of a
\PB initially being in a particular flavor state and decaying after
time \dectime
in the same flavor state ($q_f =+1$) or in the opposite flavor state ($q_f =-1$) 
is
\begin{equation}
  P(\dectime, q_f | \tauBz, \dmd)
  = \frac{e^{-\abs{\dectime} / \tauBz}}{4\tauBz} \qty[1 + q_f \cosdmd].
  \label{eqn:single_decay_prob}
\end{equation}
By measuring the distribution of \dectime and $q_f$, we determine both $\tauBz$ 
and \dmd. 
In each event, 
we fully reconstruct the ``signal-side'' \PB (\PBsig)  via 
\HepProcess{\PBzero\to\PDorDstarminus \pi^+} decays, identifying its flavor via the pion charge, as the contribution
from \HepProcess{\APBzero\to\PDorDstarminus \pi^+} decays is of the order of $10^{-4}$~\cite{Beneke:2000ry, LHCb:2018zap, BaBar:2006slj, Belle:2011dhx} 
and hence can be neglected here.
Throughout this paper, 
charge-conjugate modes are implicitly included unless stated otherwise.

We use a flavor-tagging algorithm  to
determine the flavor of the other, or ``tag-side'', \PB meson (\PBtag)
when it decays~\cite{Belle-II:2021zvj}.
As the \PB mesons are produced in a quantum-entangled state,
the flavor of \PBtag when it decays identifies (or tags) the flavor of 
\PBsig at that instant~\cite{Bigi:1981qs,Oddone:1987up}.
From that time onwards, 
the signal-side \PB freely oscillates in  flavor. 
The variable \dectime is the difference between the proper decay times of the \PBsig and
\PBtag. Equation~\ref{eqn:single_decay_prob} also applies when 
\PBsig decays first, \textit{i.e.}, for negative \dectime.

At SuperKEKB~\cite{Akai:2018mbz}, the \PUpsilonFourS is produced with a Lorentz boost in
the laboratory frame of $\beta\gamma = 0.28$. 
Since the \PB mesons are nearly at rest in the \PUpsilonFourS rest frame, 
their momenta are mostly determined by the \PUpsilonFourS boost, resulting
in a mean displacement between the \PBsig 
and \PBtag decay positions of the order of \SI{100}{\micro m} along the boost 
direction.
By measuring the
relative displacement, and knowing the \PUpsilonFourS boost, 
we determine \dectime.  To measure \tauBz and
\dmd, we fit Eq.~(\ref{eqn:single_decay_prob}), modified to account
for the \PBtag decay probability and detection effects, to the
background-subtracted \dectime distribution.

The \belleii detector consists of subsystems arranged cylindrically
around the interaction region~\cite{Belle-II:2010dht}.
The $z$ axis of the laboratory frame
is defined as the symmetry axis of the cylinder, and the
positive direction is approximately given by the electron-beam direction,
which is the beam with higher energy. 
The polar angle $\theta$, as well as the longitudinal and  transverse 
directions, are defined with
respect to the $+z$ axis.
Charged-particle trajectories (tracks) are reconstructed by a two-layer silicon-pixel
detector~(PXD) surrounded by a four-layer double-sided silicon-strip
detector~(SVD) and a 56-layer central drift chamber~(CDC). 
When the
data analyzed here were collected, only one sixth of the second PXD layer
was installed.
A quartz-based Cherenkov counter measuring the Cherenkov photon time-of-propagation
is used to identify hadrons in the central region, 
and an aerogel-based ring-imaging Cherenkov counter is used to identify hadrons 
in the forward end-cap region. An 
electromagnetic calorimeter~(ECL) is used to reconstruct photons and 
to provide information for
particle identification, in particular, to distinguish electrons from
other charged particles. 
All subsystems up to the ECL are located within 
an axially uniform \SI{1.5}{T} magnetic field provided by 
a superconducting solenoid. 
A subsystem dedicated to identifying \PKlong mesons
and muons is the outermost part of the detector.

The data is processed with the \belleii analysis software
framework~\cite{Kuhr:2018lps} using the track reconstruction
algorithm described in Ref.~\cite{BelleIITrackingGroup:2020hpx}.
We use Monte Carlo (MC) simulation to optimize selection criteria, 
determine shapes of probability density functions (PDFs), and study sources of background.
 We use
\texttt{KKMC}~\cite{Jadach:1999vf} to generate
\HepProcess{\APelectron\Pelectron\to\Pquark\APquark}, 
where $q$ indicates a $u$, $d$, $c$, or $s$ quark,
\texttt{PYTHIA8}~\cite{Sjostrand:2014zea} to simulate hadronization,
\texttt{EVTGEN}~\cite{Lange:2001uf} to simulate decays of hadrons, and
\texttt{GEANT4}~\cite{GEANT4:2002zbu} to model detector response. Our
simulation includes beam-induced backgrounds~\cite{Liptak:2021tog}.
We  optimize and fix our selection criteria
using simulated data before examining the experimental data.

We reconstruct  \HepProcess{\PBzero\to\PDstarminus\Ppiplus} and 
\HepProcess{\PBzero\to\PDminus\Ppiplus} decays by first reconstructing 
\PD
mesons via \HepProcess{\PDminus\to\PKplus\Ppiminus\Ppiminus},
\HepProcess{\APDzero\to\PKplus\Ppiminus},
\HepProcess{\APDzero\to\PKplus\Ppiminus\Ppizero}, and
\HepProcess{\APDzero\to\PKplus\Ppiminus\Ppiplus\Ppiminus} decays.
We then  reconstruct \PDstarminus mesons 
in their decay to a $\APDzero\Ppiminus$ final state,
where the pion is referred to as the ``slow pion''---one with
low momentum in the \PUpsilonFourS rest frame. 
 Finally, we combine a \PDminus or \PDstarminus candidate with a 
 charged particle identified as a pion to form  the \PBzero candidate.

We require that tracks originate from the interaction region  and have 
at least six measurement points (hits) in the SVD or twenty hits in the CDC. 
Each track must have a 
distance-of-closest-approach 
to the interaction point of less than~\SI{3}{cm} 
along the $z$ axis and less than~\SI{0.5}{cm} in the plane
transverse to it, and have a polar angle in the CDC acceptance range~$[17^{\circ},
  150^{\circ}]$.
These requirements 
reduce backgrounds with poorly reconstructed tracks and tracks from beam background.

Photon candidates are identified as localized energy deposits in the ECL 
not associated with any track. 
To suppress beam-induced photons, which have different energy spectra depending on 
their momentum direction,
each photon is required to have an energy greater than $\SI{30}{MeV}$ if reconstructed in the
 central region of the calorimeter, greater than $\SI{80}{MeV}$ if reconstructed in the backward region, 
 and greater than $\SI{120}{MeV}$ if reconstructed in the forward region. 
Neutral pions are reconstructed from pairs of photon candidates that have
an angular separation of less than~$52^{\circ}$ in the lab frame and an invariant mass
in the range~$[121\,\si{MeV\naturalunits{/c^2}},142\,\si{MeV\naturalunits{/c^2}}]$.

We reconstruct \PD mesons by combining two to four particles, one of them
being identified as a $\PKplus$.
The mass of $\APDzero$ candidates must be in the range 
$[\SI{1.845}{MeV}, \SI{1.885}{MeV}]$ for $\APDzero\to K^+\pi^-$
and $\APDzero \to K^+\pi^-\pi^+\pi^-$, and in the range $[\SI{1.810}{MeV},\SI{1.895}{MeV}]$
for  $\APDzero\to K^+\pi^-\pi^0$. The mass of $\PDminus$ candidates is required 
to be in the range $[\SI{1.860}{MeV}, \SI{1.880}{MeV}]$. The mass range is looser for 
$\APDzero$ candidates, as the selection requirements placed on the  $\PDstarminus$ are sufficient
to suppress background events containing a fake $\APDzero$.

We identify negatively charged pions with momenta  below \SI{300}{MeV\naturalunits{/c}} 
in the center-of-mass frame as slow pion candidates. 
Each of these candidates  is combined with a \APDzero candidate
to form a \PDstarminus
candidate. The energy released in the \PDstarminus decay,
$m(\PDstarminus)-m(\APDzero)-m_{\pi^+}$, must be
in the range~$[4.6\,\si{MeV}, 7.0\,\si{MeV}]$. 

Each \PDorDstarminus is combined with a remaining positive particle to form a \PBsig
candidate. To remove background  from
\HepProcess{\PBzero\to\PDorDstarminus\Pleptonplus\Pnulepton} decays, we
require the particle to be identified as a pion. 
A small number of Cabibbo-suppressed \HepProcess{\PBzero\to\PDorDstarminus\PKplus} 
decays pass this requirement. Their yield is $2.7\%$ of that of \HepProcess{\PBzero\to\PDorDstarminus\Ppiplus}
decays. These decays have the same \dectime distribution 
as \HepProcess{\PBzero\to\PDorDstarminus\Ppiplus}, and we treat them as signal.

We identify \PBsig candidates using two quantities, the 
beam-constrained mass $M\sub{bc}$ and the energy difference $\Delta E$. 
These quantities are defined as
\begin{equation}
	M\sub{bc} \equiv \sqrt{E_{\text{beam}}^2 - \abs*{\vec{p}}^2} \naturalunits{c^2},
  \quad
  \Delta E \equiv E - E_{\text{beam}},
\end{equation}
where $E_{\text{beam}}$ is the beam energy, and $\vec{p}$ and $E$ are the 
reconstructed momentum and energy, respectively, of the \PBsig candidate. All quantities 
are calculated in the 
the \HepProcess{\APelectron\Pelectron} center-of-mass frame. 
We calculate $E$ assuming that the  track directly from \PBsig is a pion.
We require that $M\sub{bc}$ be greater than~\SI{5.27}{GeV\naturalunits{/c^2}} and
that $\Delta E$ be in the range~$[-0.10\,\si{GeV}, 0.25\,\si{GeV}]$.
The $\Delta E$ range is  asymmetric, \textit{i.e.}, shorter on the lower side, 
to reduce backgrounds from $B$ decays with missing daughters.

We determine the \PBtag vertex and flavor using the remaining tracks
in the event.  Such tracks are required to have at least one hit in each of
the PXD, SVD, and CDC and have a reconstructed momentum greater
than \SI{50}{MeV\naturalunits{/c}}. Each track must also originate
from the $\APelectron\Pelectron$ interaction point according to the
same criteria as used to select \PBsig candidates. 
We require that the \PBtag decay includes at least one
charged particle. 
The \PBtag momentum is taken to be opposite that of the \PBsig candidate in 
the center-of-mass frame.

To determine the \PBsig decay vertex, we fit its decay chain with
the \texttt{TreeFit} algorithm~\cite{Amoraal:2012qn,krohn}. To determine the
 \PBtag decay vertex, we fit its decay products with the
\texttt{Rave} adaptive algorithm~\cite{Waltenberger:2008zza}, which
accounts for our lack of knowledge of the decay chain by reducing the
impact of tracks displaced by potential intermediate \PD decays. 
The decay vertex position is adjusted  such that the direction of each \PBzero, 
as determined from its decay vertex and the $\APelectron\Pelectron$
interaction point~\cite{Dey:2020dsr}, is parallel to its momentum vector.
The IR is measured from
\HepProcess{\APelectron\Pelectron\to\APmuon\Pmuon} events.
Charged \PD
candidates must have positive flight distances. 
We require that both vertex
fits converge, and that the uncertainty on the decay time,
\dectimeunc, as calculated from the fitted vertex positions,
be less than~\SI{2}{ps}. 
These vertex quality requirements retain
approximately $90\%$ of signal events.

The efficiency to reconstruct a  $\PBsig\PBtag$ pair with 
 \HepProcess{\PBsig\to\PDminus\Ppiplus} 
is~\num{34}\%. For \HepProcess{\PBsig\to\PDstarminus\Ppiplus} with
\HepProcess{\APDzero\to\PKplus\Ppiminus}, it is~\num{35}\%; with
\HepProcess{\APDzero\to\PKplus\Ppiminus\Ppizero}, it is~\num{15}\%;
and with \HepProcess{\APDzero\to\PKplus\Ppiminus\Ppiplus\Ppiminus}, it
is~\num{25}\%. 
In \num{2.2}\% of selected events, there is more than one $\PBsig$ candidate. 
We retain all such candidates for further analysis.

The main sources of background are misreconstructed \HepProcess{\PUpsilonFourS\to\PB\APB} events 
and nonresonant $e^+e^-\to\Pquark\APquark$  events. 
To distinguish between signal and $\Pquark\APquark$, we train two multivariate 
classifiers~\cite{xgboost}: one for \HepProcess{\PBzero\to\PDminus\Ppiplus} decays and one for 
\HepProcess{\PBzero\to\PDstarminus\Ppiplus} decays. The classifiers exploit 
the difference in event topologies and use as input the following 
quantities: Fox-Wolfram moments~\cite{fw} and an extension thereof~\cite{Belle:ksfw};
``cone'' variables developed by the CLEO collaboration~\cite{CLEO:1995rok}; 
the angle between the thrust axes of the two \PB mesons~\cite{BaBar:2014omp}; 
and the event sphericity~\cite{Parisi:1978eg}.  %~\cite{Belle-II:2018jsg}. 
The classifiers are trained and tested using simulated data. 
In addition to determining the flavor of each \PBtag, the 
flavor-tagging algorithms return a tag-quality variable $r$,
which ranges from $0$ for no flavor information to $+1$ for unambiguous flavor 
assignment.
From the \PBtag and \PBsig flavors,  we
determine the relative flavor $q_f$.
The data is divided into seven subsamples,
depending on the $r$ value: $[0.0, 0.10]$, 
$[0.10,0.25]$, $[0.25,0.45]$, $[0.45,0.60]$, $[0.60,0.725]$, $[0.725,0.875]$,
and $[0.875,1.0]$. This division enhances the statistical precision of the $\dmd$
measurement.

We determine the signal yield by performing an unbinned, extended maximum-likelihood fit 
to the distributions of $\Delta E$ and the multivariate-classifier output $C$.
The fit is performed
separately 
for each $r$ interval and determines the yield of signal events and $\PB\APB$ and 
$\Pquark\APquark$ background events. As the fit observables $\Delta E$ and $C$ are 
found to have negligible correlation, the PDFs ($P$) for these variables
are taken to factorize:
 $P(\Delta E, C) = P(\Delta E)\cdot P(C)$. 
All PDFs are determined separately for each $r$ interval; however, some of the parameters 
(as noted below) are taken to be common among the $r$ intervals.

The $\Delta E$ PDF for signal is modeled as the sum of two double-sided Crystal Ball 
functions~\cite{Gaiser:1982yw}: one for 
\HepProcess{\PBzero\to\PDorDstarminus\Ppiplus} decays and one for
\HepProcess{\PBzero\to\PDorDstarminus\PKplus} decays. The shape parameters of these functions, 
as well as the ratio between their normalizations, are fixed to values obtained from  
simulation. To account for differences between data and simulation, we introduce two additional 
free parameters: a shift of the mean values of the functions, and a scale factor for their widths.
These parameters are taken to be common among the $r$ intervals. The $\Delta E$ PDF for $\PB\APB$ background
is a fourth-order polynomial, and the $\Delta E$ PDF for $\Pquark\APquark$ background 
is an exponential function. 
All parameters of the polynomial are fixed to values obtained from simulation,
while the slope of the exponential function is free to vary.

The $C$ PDFs for signal and background are taken to be Johnson $S_U$ functions~\cite{johnson}. 
The Johnson functions across different $r$ intervals have independent
mode, standard deviation, skewness, and kurtosis parameters, all determined 
from simulation.
We  introduce four free parameters to account for differences between data and simulation 
that are common across all $r$ intervals: one offset for the modes
and one scale for the widths for all
$\Pquark\APquark$-background distributions; and similarly one offset
and one scale common to all signal and $\PB\APB$-background
distributions.

\begin{figure}[htbp]
  \centering
  
  \includegraphics[width=\columnwidth]{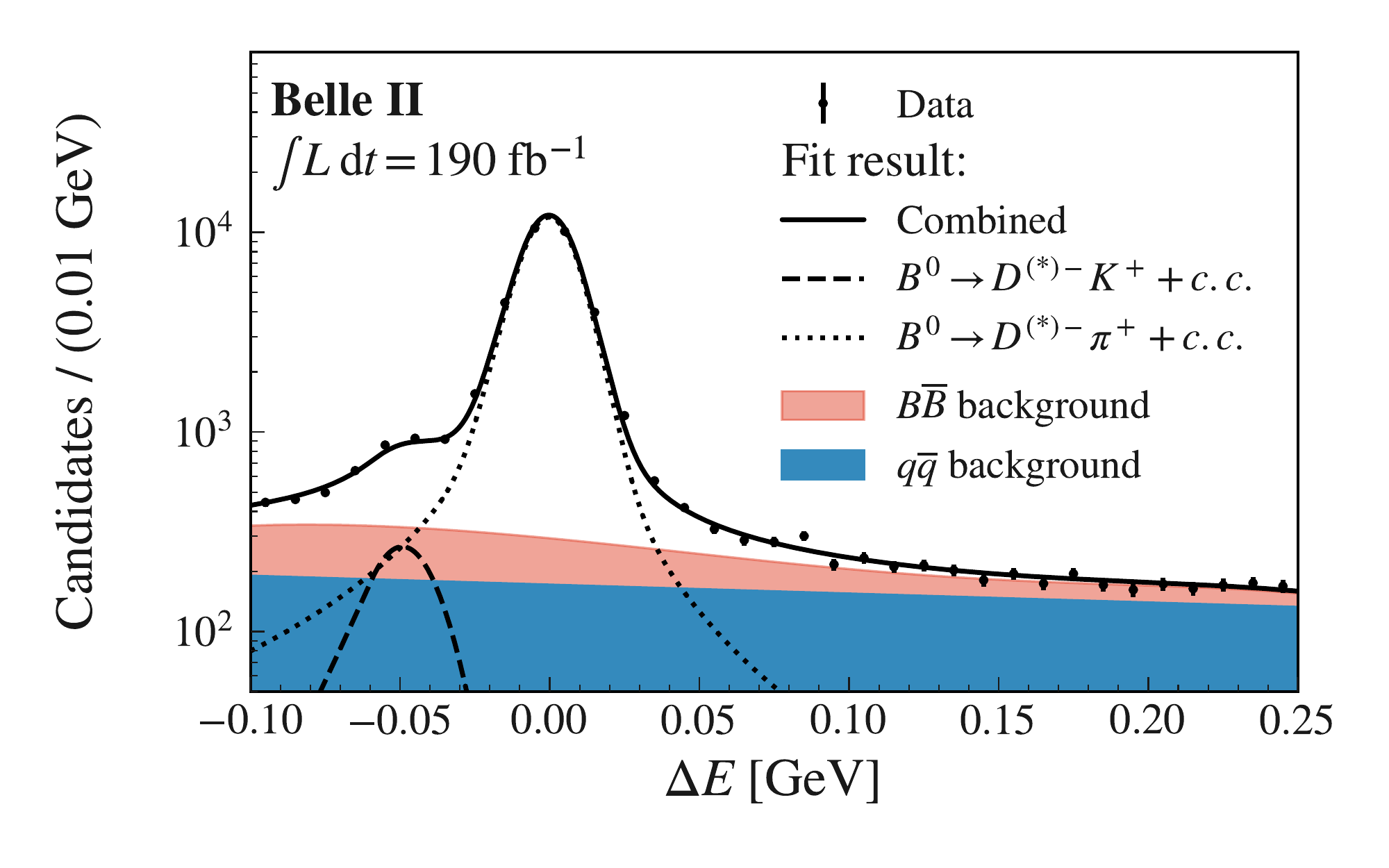}\\
  \includegraphics[width=\columnwidth]{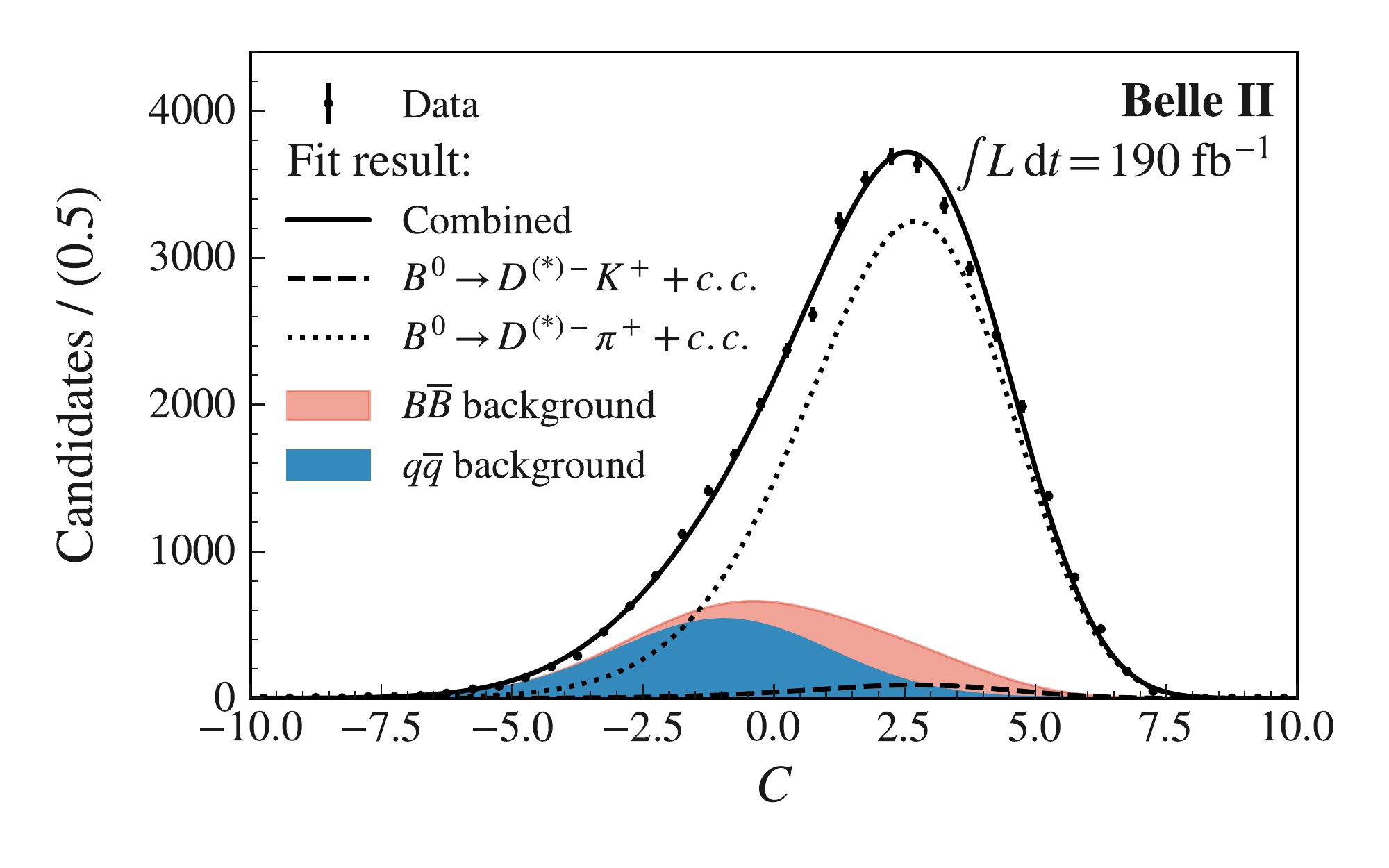}
  
  \caption{Distributions of $\Delta E$~(top) and $C$~(bottom)  in
    data~(points) and the fit model~(curves and stacked shaded regions).
  }
  
  \label{fig:DeltaE_C}
\end{figure}

We simultaneously fit to data in all seven $r$ intervals. The fit has a total of 28 free parameters: 
three yields for each of the $r$ intervals, six scale or shift factors, and the slope of the exponential function 
used for the $\Delta E$ PDF of the $\Pquark\APquark$ background. The distributions of $\Delta E$ and $C$
summed over all $r$ intervals, 
along with projections of the fit results, are shown in Fig.~\ref{fig:DeltaE_C}. 
The resulting yields are~\num{33317 +- 203}~signal events, \num{2814 +- 150}~$\PB\APB$-background
events, and~\num{5594 +- 125}~$\Pquark\APquark$-background events.

%%%%%%%%%%%%%%%%%%%%%%%%%%%%%%%%%%%%%%%%%%%%%%%%%%
%%%%%%%%%%%%%%%%%%%%%%%%%%%%%%%%%%%%%%%%%%%%%%%%%%

Using 
sWeights~\cite{Pivk:2004ty, Dembinski:2021kim} computed with the per-candidate signal fractions
obtained from the fit to $\Delta E$ and $C$,
we statistically subtract background contributions to the \dectime and \dectimeunc
distributions. In this manner,  we need not parametrize background
distributions when fitting for \tauBz and \dmd.

We measure the lifetime \tauBz and oscillation frequency \dmd by fitting the
background-subtracted \dectime and \dectimeunc distributions.
The probability density to observe both \PBsig and \PBtag decays is obtained 
from eq.~(\ref{eqn:single_decay_prob}) by including the probability for 
\PBtag to decay and the probability of mistagging its flavor,
\begin{alignat}{1}
  P(\dectime, \avgdectime, q_f, r | \tauBz, \dmd)
  &= \frac{e^{-2\avgdectime/\tauBz}}{\tauBz[2]} \nonumber
  \times \{1 + \\& q_f [1 - 2w(r)] \cosdmd\},
  \label{eqn:double_decay_prob}
\end{alignat}
where \avgdectime is the average of the \PBsig and \PBtag proper decay times,
and $w(r)$ is the probability of the \PBtag flavor being incorrectly
assigned. The latter is parametrized with a single value for each $r$
interval and is assumed to be independent of the \PBtag flavor.

The decay-time difference \dectime can be expressed as
\begin{equation}
  \dectime = \measdectime - 2\frac{\beta_{\PB}}{\beta} \, \avgdectime \cos\theta,
\end{equation}
where $\measdectime \equiv \declength / (\beta\gamma\gamma_{\PB})$.
In this expression,
\declength is the displacement of the \PBsig vertex from that of 
\PBtag, 
$\beta$ is the
velocity of the \PUpsilonFourS in the lab frame (with $\gamma = (1-\beta^2)^{-\frac{1}{2}}$), 
$\beta_B$ is the velocity of a \PB in the \PUpsilonFourS rest frame,
and $\theta$ is the polar angle
 of the \PBsig direction in the \PUpsilonFourS rest frame.
We integrate out the dependence of
eq.~(\ref{eqn:double_decay_prob}) on \avgdectime and $\theta$, accounting for
the  angular distribution in
\HepProcess{\APelectron\Pelectron\to\PUpsilonFourS\to\PB\APB},
  $P_{\theta}(\cos\theta) = (3/4) (1 - \cos^2\theta)$.

To account for resolution and bias in measuring \declength, we
convolve eq.~(\ref{eqn:double_decay_prob}) with an empirical response function,
which is modeled as a linear combination
 of three components:
\begin{equation}
	\label{eqn:resp_fun}
        \begin{aligned}
            \resfunc(\dectimeoffset|\dectimeunc) &= (1-f_{\tailsub}-f_{\OLsub}) G(\dectimeoffset| m_{\gausfunc}\dectimeunc,s_{\gausfunc}\dectimeunc)\\ 
            &+f_{\tailsub}(\dectimeunc)\resfunc_{\tailsub}(\dectimeoffset| m_{\tailsub}\dectimeunc,s_{\tailsub}\dectimeunc, k/\dectimeunc, f_{>}, f_{<})\\ 
            &+f_{\OLsub} \gausfunc(\dectimeoffset| 0, \sigma_{0}),
        \end{aligned}
\end{equation}
where $\dectimeoffset \equiv (\declength - \declength\sub{true}) /
(\beta\gamma \gamma_{\PB})$ and $\declength\sub{true}$ is the true value
of~\declength. 
The first component is a Gaussian distribution with mean and standard
deviation proportional to the per-candidate \dectimeunc; this component
accounts for $70\%$ of candidates. The second component is
a weighted sum of a Gaussian distribution and two
exponentially modified Gaussian functions, corresponding to a 
Gaussian convolved with an exponential distribution, 
    \begin{equation}
        \begin{aligned}
        \resfunc_{\tailsub}(x|\mu,\sigma,\kappa, f_{<}, f_{<})&= (1-f_<-f_>) G(x|\mu,\sigma)\\
        &+f_{<} G(x|\mu,\sigma)\otimes \kappa\exp_{<}(\kappa x)\\
        &+f_{>} G(x|\mu,\sigma)\otimes \kappa\exp_{>}(-\kappa x),
        \end{aligned}
    \end{equation}
    where $\exp_{>}(-\kappa x)=\exp(-\kappa x)$ if $x>0$ and $\exp_{>}(-\kappa x)=0$ otherwise, and
    similarly for $\exp_{<}(\kappa x)$. 
The exponential tails  account for poorly determined
\PBtag vertices due to intermediate
charm mesons yielding displaced secondary vertices. 
The fraction $f_{\tailsub}$ is zero at low values of 
\dectimeunc and reaches a plateau of $0.2$ at approximately $\dectimeunc=\SI{25}{\pico\second}$. 
This is modeled using three parameters:
the maximal tail fraction $f_{\tailsub}^{\text{max}}$ at its plateau,
a threshold parameter describing the \dectimeunc value at 
which the tail fraction becomes nonzero, and a slope parameter describing 
how fast the tail fraction reaches its plateau. 
The third component 
has a large width, $\sigma_0=\SI{200}{\pico\second}$,
to account for the $\mathcal{O}({10^{-3}})$ fraction of outlying
poorly reconstructed vertices.

\begin{figure}[!htbp]
  \centering
  \vspace{0.1cm}

  \includegraphics[width=\columnwidth]{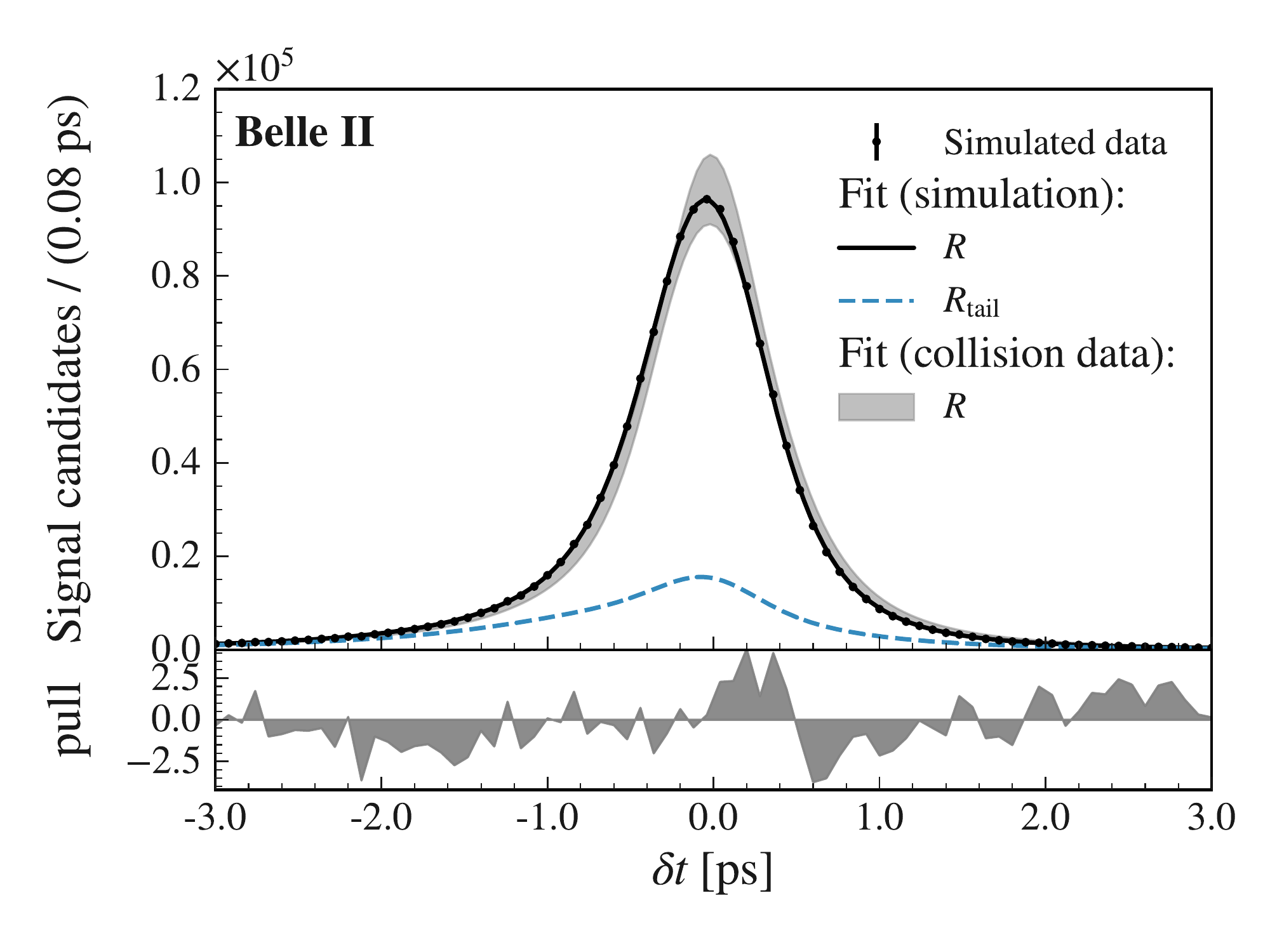}

  \caption{Top: Distribution of \dectimeoffset  in simulated data~(points) and
    distribution modeled by the response function
    from the fit to the simulated data~(curves) and
    from the fit to the experimental data~(shaded). The shaded area accounts 
    for the statistical and systematic uncertainties on the parameters
    of the response function.
    Bottom: distribution of the pull, defined as the difference between the event count 
    in each bin and its value predicted by the fit, divided by the Poisson uncertainty.
  }

  \label{fig:detector_response}
\end{figure}

Equation~(\ref{eqn:resp_fun}) is the simplest model found to satisfactorily 
describe the
\dectimeoffset distribution of simulated events. We fix 
$\sigma_0$, as well as  
$k$, $f_>$, $f_<$, and the $f_{\tailsub}$ slope and threshold parameters, 
to values determined from
a fit to simulated data. Figure~\ref{fig:detector_response} shows the
\dectimeoffset distribution of simulated data and the distribution of the fitted model. %We determine $f_{\tailsub}$, the
The parameter $f_{\tailsub}^{\text{max}}$, as well as the scaling factors relating 
the modes and standard deviations of $G$ and $\resfunc_{\tailsub}$ to \dectimeunc
--- $m_{\gausfunc}$, $s_{\gausfunc}$,  $m_{\tailsub}$ and $s_{\tailsub}$ ---
are free to vary in the fit to data.

%%%%%%%%%%%%%%%%%%%%%%%%%%%%%%%%%%%%%%%%%%%%%%%%%%
%%%%%%%%%%%%%%%%%%%%%%%%%%%%%%%%%%%%%%%%%%%%%%%%%%

After integrating over $\cos\theta$ and \avgdectime and convolving with
$\resfunc(\dectimeoffset)$, the \measdectime distribution of $\PB$ meson pairs 
is 
\begin{widetext}
  \begin{equation}
    P(\measdectime, \dectimeunc, q_f, r | \tauBz, \dmd)
    = P(\dectimeunc|q_f, r)
    \!\!\int\!\!
    P(\dectime - \dectimeoffset, \avgdectime, q_f, r | \tauBz, \dmd)
    \hspace{0.1em} P_\theta(\cos\theta) \, \resfunc(\dectimeoffset|\dectimeunc)
    \hspace{0.1em} \dd\dectimeoffset \, \dd\!\cos\theta \, \dd\avgdectime,
    \label{eqn:marginalized_prob}
    \end{equation}
\end{widetext}
where $P(\dectimeunc|q_f, r)$ is the probability to observe \dectimeunc for 
a given value of $q_f$ and $r$, modelled using histogram templates:
one for each $r$ interval and
value of $q_f$~($14$ in total), taken from the data.
The sWeights computed using the fit to $\Delta E$ and $C$ are used to 
statistically subtract the background contribution to the \dectimeunc histograms. 
We fit for \tauBz and \dmd by maximizing
\begin{equation}
  \sum_{i} s^i \ln P(\measdectime^i,\dectimeunc^i, q_f^i, r^i | \tauBz, \dmd),
  \label{eq:dll}
\end{equation}
where the sum runs over all $\PBsig\PBtag$ candidate pairs and $s^i$ 
is the sWeight of a pair. 
 Fourteen parameters are free in the fit: \tauBz and
\dmd; seven values of $w$, one for each $r$ interval; and the five
free parameters of the response function.

We calculate the statistical uncertainties using
$1000$  bootstrapped~\cite{bootstrap} samples obtained from the 
data. 
For each sample, we repeat the determination of the sWeights and the fit for
\tauBz and \dmd. In this way, the spread of fitted \tauBz and \dmd values
account for the statistical fluctuations of the signal and background fractions.
We test this analysis method with independent
simulated data. 
When tested on simulated data, our fitting procedure determines \tauBz
with a small systematic bias of \SI{0.004\pm0.002}{ps} and \dmd 
with no significant bias, $\SI{0.000\pm 0.001}{ps^{-1}}$.
We assign the central value of the bias on \tauBz as a systematic
uncertainty. We assign the uncertainty on the bias on \dmd, arising 
from the size of the simulated data, as a systematic uncertainty.

\begin{figure}[htbp]
  \centering

  \includegraphics[width=\columnwidth]{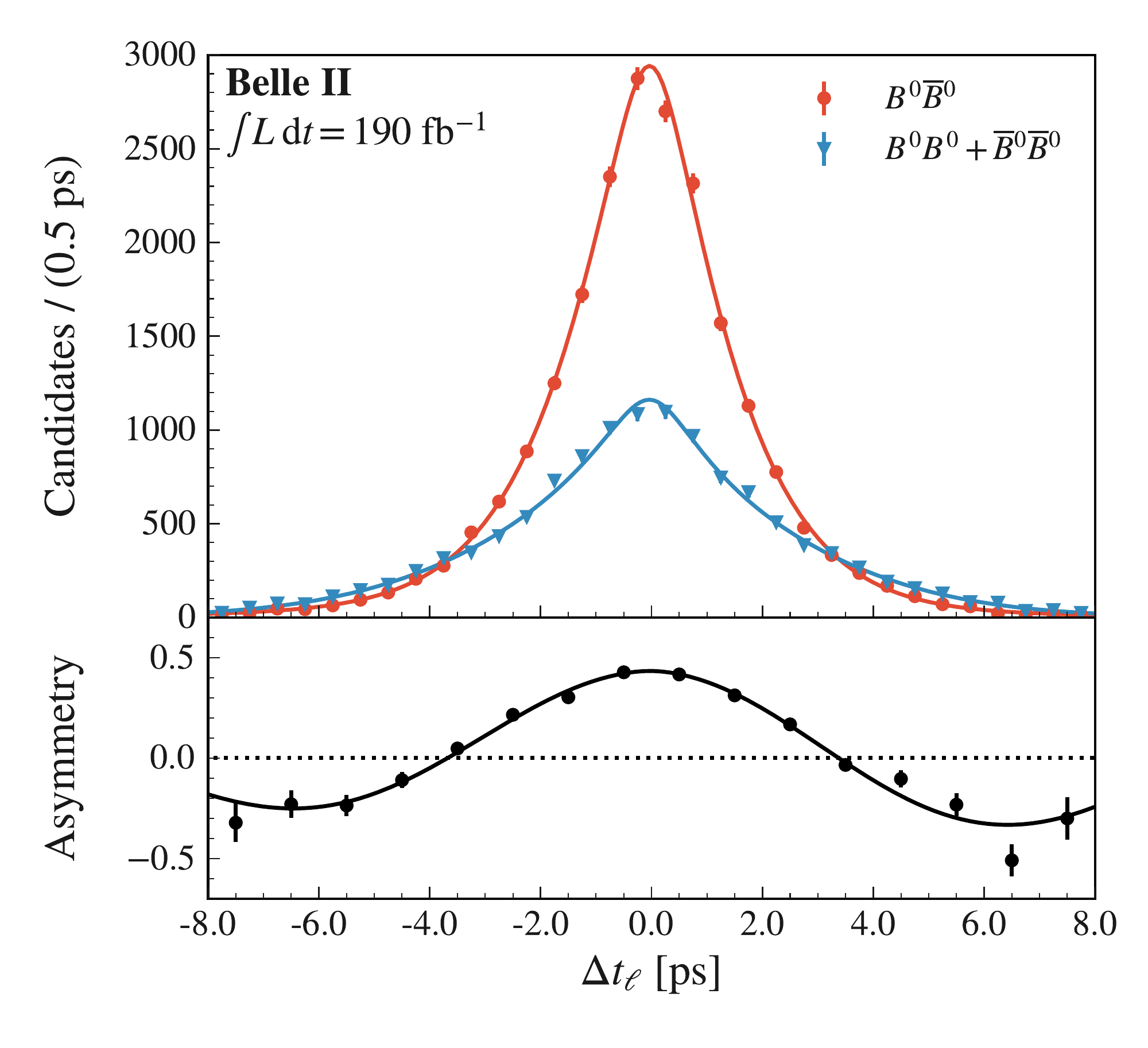}

  \caption{Distribution of \measdectime in data~(points) and the fit
    model~(lines) for
    opposite-flavor candidate pairs~(red) and same-flavor pairs~(blue) 
    and their asymmetry~(black).}
  
  \label{fig:fit_results}
\end{figure}

The \measdectime distributions of both opposite-flavor and same-flavor 
$\PB$-meson  pairs are shown in Fig.~\ref{fig:fit_results} for all $r$ intervals
combined, along with 
projections of the fit result. 
We also check that the fit quality is good in each individual $r$ interval.
The figure shows the $\measdectime$-dependent
yield  asymmetry 
between the two samples, defined as the difference between the number 
of opposite-flavor pairs and same-flavor pairs divided by their sum.
The fit results and statistical 
uncertainties for \tauBz and \dmd are
\SI{1.499 +- 0.013}{ps} and
\SI{0.516 +- 0.008}{ps^{-1}\naturalunits{\hbar/c^2}},
with a $-29\%$ statistical correlation factor between them.

%%%%%%%%%%%%%%%%%%%%%%%%%%%%%%%%%%%%%%%%%%%%%%%%%%

There are several sources of systematic uncertainty; 
these are listed in Tab.~\ref{tab:syst_uncs} and described below.
The dominant systematic uncertainty is due to potential discrepancies
between the assumed values (fixed in the fit) of the response-function parameters and the
true values in the data. For each fixed parameter, we repeat the fit
with the parameter allowed to vary. 
We add all the resulting changes in the result in quadrature and 
include this value as a systematic uncertainty.

%%%%%%%%%%%%%%%%%%%%%%%%%%%%%%%%%%%%%%%%%%%%%%%%%%
\begin{table}[htbp]
  \centering
  
  \caption{Systematic uncertainties.}

  %% Set precision to 3 decimal places
  \sisetup{round-mode=places,round-precision=3}

  \begin{tabular}{lSS}
    
    %\toprule
    %\toprule
    \hline\hline
    
    Source
    & {\tauBz~[\si{ps}]}
    & {\dmd~[\si{ps^{-1} \naturalunits{\hbar/c^2}}]}
    \\
    
    %\midrule
    \hline

    %%%%%%%%%%%%%%%%%%%%%%%%%
    %% 1
    Fixed response-function parameters
    & 0.0063
    & 0.0028
    \\

    %%%%%%%%%%%%%%%%%%%%%%%%%
    %% 2
    Analysis bias
    & 0.0043
    & 0.0011
    \\

    %%%%%%%%%%%%%%%%%%%%%%%%%
    %% 3
    Detector alignment
    & 0.0027
    & 0.0024
    \\

    %%%%%%%%%%%%%%%%%%%%%%%%%
    %% 4
    Interaction-region precision
    & 0.0021
    & 0.0014
    \\

    %%%%%%%%%%%%%%%%%%%%%%%%%
    %% 5
    $C$-Distribution modeling
    & 0.0000
    & 0.0014
    \\

    %%%%%%%%%%%%%%%%%%%%%%%%%
    %% 6
    \dectimeunc-Distribution modeling
    & 0.0005
    & 0.0010
    \\

    %%%%%%%%%%%%%%%%%%%%%%%%%
    %% 7
    Correlations of $\Delta E$ or $C$ and $\measdectime$
    & 0.0007
    & 0.0003
    \\

    \hline

    %%%%%%%%%%%%%%%%%%%%%%%%%
    Total systematic uncertainty
    & 0.0084 
    & 0.0045 
    \\ 

    \hline

    %%%%%%%%%%%%%%%%%%%%%%%%%
    Statistical uncertainty
    & 0.0130
    & 0.0079
    \\

    \hline
    \hline
  \end{tabular}

  \vspace\baselineskip

  \label{tab:syst_uncs}
\end{table}

Possible misalignment of the tracking detector can bias our
results~\cite{Bilka:2021rqj}. To estimate this effect, we reconstruct
simulated signal events with several  misalignment
scenarios. Two scenarios are extracted from collision data using day-by-day
variations of the detector alignment.
Two additional scenarios correspond to misalignments
remaining after applying the alignment procedure to dedicated simulated data.
We repeat the analysis for each scenario and assign the
largest changes in the results as systematic uncertainties.

Because we adjust the $\PBsig$ decay vertex position so that the vector connecting 
the IR and decay vertex is parallel to the $\PBsig$ momentum,
the precision to which we know the IR affects our determination of
\declength. We repeat our analysis on simulated data in which we shift,
rotate, and rescale the IR within its measured
uncertainties and assign the changes in the results as systematic
uncertainties. We perform an analogous check with changes to
$\sqrt{s}$ and the magnitude and direction of the boost vector and
find that the results change negligibly.

%%%%%%%%%%%%%%%%%%%%%%%%%
We estimate systematic uncertainties due to mismodeling the $C$ distribution,
including possible correlation with $\Delta E$,
 from the changes in the results observed when fitting to the $\Delta E$ distribution
 only. In that case, 
 the  $\PB\APB$-background fraction is fixed
to the value in simulated data. The result for \tauBz changes
negligibly, but a systematic uncertainty is included for \dmd. To
check for dependence of the results on the $\Delta E$ model
for the $\Pquark\APquark$ and $\PB\APB$ backgrounds, we
repeat the analysis with each model replaced by a second-order
polynomial with all parameters free in the fit. The polynomial 
parameters are  common to all $r$ intervals. 
The results change negligibly.

%%%%%%%%%%%%%%%%%%%%%%%%%

%%%%%%%%%%%%%%%%%%%%%%%%%
%% Syst 7) \dectimeunc model
To check for dependence of the results on the
\dectimeunc model, we repeat the fit 
with several alternative binning choices for their templates,
and also replacing templates with analytical functions.
We assign the largest changes in the results as
systematic uncertainties.

%%%%%%%%%%%%%%%%%%%%%%%%%
%% Syst 10) Kaonic-decay fraction
We investigate the impact of fixing the yield of 
\HepProcess{\PBzero\to\PDorDstarminus\PKplus} decays
relative to \HepProcess{\PBzero\to\PDorDstarminus\Ppiplus}
by repeating the analysis with alternative choices of the 
\HepProcess{\PBzero\to\PDorDstarminus\PKplus} fraction,  
corresponding to varying the branching fractions and 
relevant hadron identification efficiencies by their known uncertainties~\cite{PDG}.
The results change negligibly.

%%%%%%%%%%%%%%%%%%%%%%%%%
%% cross check of \measdectime dependence
To check if potential correlations of $\Delta E$ or $C$ with
\measdectime affect our results, we repeat the analysis with sWeights
calculated independently for two subgroups of candidate pairs, defined by
the sign of \measdectime. Likewise, we repeat the analysis for two subgroups defined by
whether $\abs*{\measdectime}$ is greater or less than \SI{1.150}{ps}. In both
cases, the results change mildly and we assign the larger of these two 
changes as systematic uncertainties.

%%%%%%%%%%%%%%%%%%%%%%%%%
%% Syst 8) Global momentum scale
The global momentum scale of the  Belle II tracking detector 
is calibrated to a relative
precision of better than $0.1\%$, and the global length scale to a
precision of better than $0.01\%$. Neither significantly impacts our
results.

%%%%%%%%%%%%%%%%%%%%%%%%%
%% cross check of data-taking period or D vs D*
We further check our analysis by repeating it on subsets 
of the data divided by data-taking period or by whether the charm
meson in the \PBsig decay is \PDminus or $\PD^{*-}$. The results are all
statistically consistent with each other and with our overall results.

%%%%%%%%%%%%%%%%%%%%%%%%%
%% Summary
In summary, we measure the \PBzero lifetime and flavor-oscillation
frequency using \HepProcess{\PBzero\to\PD^{(*)-}\Ppiplus} decays 
reconstructed in data
 collected from $\APelectron\Pelectron$ collisions at the \PUpsilonFourS resonance 
 and corresponding to an integrated luminosity of \SI{190}{fb^{-1}}.
The results are
\begin{alignat}{1}
  \tauBz &= \tauBzResult\\
  \dmd   &= \dmdResult.
\end{alignat}
The results agree with previous measurements 
and have very similar systematic uncertainties as compared  
to results from the Belle and Babar collaborations~\cite{Belle:2004hwe,BaBar:2005laz}.
They demonstrate a good understanding of the Belle~II detector and provide a 
strong foundation for future time-dependent measurements.

\begin{acknowledgments}
% Policy from October 20, 2022
This work, based on data collected using the Belle II detector, which was built and commissioned prior to March 2019, was supported by
%Armenia
Science Committee of the Republic of Armenia Grant No.~20TTCG-1C010;
%Australia
Australian Research Council and research Grants
No.~DE220100462,
No.~DP180102629,
No.~DP170102389,
No.~DP170102204,
No.~DP150103061,
No.~FT130100303,
No.~FT130100018,
and
No.~FT120100745;
%Austria
Austrian Federal Ministry of Education, Science and Research,
Austrian Science Fund
No.~P~31361-N36
and
No.~J4625-N,
and
Horizon 2020 ERC Starting Grant No.~947006 ``InterLeptons'';
%Canada
Natural Sciences and Engineering Research Council of Canada, Compute Canada and CANARIE;
%China
Chinese Academy of Sciences and research Grant No.~QYZDJ-SSW-SLH011,
National Natural Science Foundation of China and research Grants
No.~11521505,
No.~11575017,
No.~11675166,
No.~11761141009,
No.~11705209,
and
No.~11975076,
LiaoNing Revitalization Talents Program under Contract No.~XLYC1807135,
Shanghai Pujiang Program under Grant No.~18PJ1401000,
Shandong Provincial Natural Science Foundation Project~ZR2022JQ02,
and the CAS Center for Excellence in Particle Physics (CCEPP);
%Czech Republic
the Ministry of Education, Youth, and Sports of the Czech Republic under Contract No.~LTT17020 and
Charles University Grant No.~SVV 260448 and
the Czech Science Foundation Grant No.~22-18469S;
%EU
European Research Council, Seventh Framework PIEF-GA-2013-622527,
Horizon 2020 ERC-Advanced Grants No.~267104 and No.~884719,
Horizon 2020 ERC-Consolidator Grant No.~819127,
Horizon 2020 Marie Sklodowska-Curie Grant Agreement No.~700525 "NIOBE"
and
No.~101026516,
and
Horizon 2020 Marie Sklodowska-Curie RISE project JENNIFER2 Grant Agreement No.~822070 (European grants);
%France
L'Institut National de Physique Nucl\'{e}aire et de Physique des Particules (IN2P3) du CNRS (France);
%Germany
BMBF, DFG, HGF, MPG, and AvH Foundation (Germany);
%India
Department of Atomic Energy under Project Identification No.~RTI 4002 and Department of Science and Technology (India);
%Israel
Israel Science Foundation Grant No.~2476/17,
U.S.-Israel Binational Science Foundation Grant No.~2016113, and
Israel Ministry of Science Grant No.~3-16543;
%Italy
Istituto Nazionale di Fisica Nucleare and the research grants BELLE2;
%Japan
Japan Society for the Promotion of Science, Grant-in-Aid for Scientific Research Grants
No.~16H03968,
No.~16H03993,
No.~16H06492,
No.~16K05323,
No.~17H01133,
No.~17H05405,
No.~18K03621,
No.~18H03710,
No.~18H05226,
No.~19H00682, % Niigata
No.~22H00144,
No.~26220706,
and
No.~26400255,
the National Institute of Informatics, and Science Information NETwork 5 (SINET5), 
and
the Ministry of Education, Culture, Sports, Science, and Technology (MEXT) of Japan;  
%Korea
National Research Foundation (NRF) of Korea Grants
No.~2016R1\-D1A1B\-02012900,
No.~2018R1\-A2B\-3003643,
No.~2018R1\-A6A1A\-06024970,
No.~2018R1\-D1A1B\-07047294,
No.~2019R1\-I1A3A\-01058933,
No.~2022R1\-A2C\-1003993,
and
No.~RS-2022-00197659,
Radiation Science Research Institute,
Foreign Large-size Research Facility Application Supporting project,
the Global Science Experimental Data Hub Center of the Korea Institute of Science and Technology Information
and
KREONET/GLORIAD;
%Malaysia
Universiti Malaya RU grant, Akademi Sains Malaysia, and Ministry of Education Malaysia;
%Mexico
% CINVESTAV-IPN, UNAM, UAS, BUAP and CONACYT are funded under
Frontiers of Science Program Contracts
No.~FOINS-296,
No.~CB-221329,
No.~CB-236394,
No.~CB-254409,
and
No.~CB-180023, and No.~SEP-CINVESTAV research Grant No.~237 (Mexico);
%Poland
the Polish Ministry of Science and Higher Education and the National Science Center;
%Russia
the Ministry of Science and Higher Education of the Russian Federation,
Agreement No.~14.W03.31.0026, and
the HSE University Basic Research Program, Moscow;
%Saudi Arabia
University of Tabuk research Grants
No.~S-0256-1438 and No.~S-0280-1439 (Saudi Arabia);
%Slovenia
Slovenian Research Agency and research Grants
No.~J1-9124
and
No.~P1-0135;
%Spain
Agencia Estatal de Investigacion, Spain
Grant No.~RYC2020-029875-I
and
Generalitat Valenciana, Spain
Grant No.~CIDEGENT/2018/020
%Taiwan
Ministry of Science and Technology and research Grants
No.~MOST106-2112-M-002-005-MY3
and
No.~MOST107-2119-M-002-035-MY3,
and the Ministry of Education (Taiwan);
%Thailand
Thailand Center of Excellence in Physics;
%Turkey
TUBITAK ULAKBIM (Turkey);
%Ukraine
National Research Foundation of Ukraine, project No.~2020.02/0257,
and
Ministry of Education and Science of Ukraine;
%USA
the U.S. National Science Foundation and research Grants
No.~PHY-1913789 % Indiana CEEM
and
No.~PHY-2111604, % Luther
and the U.S. Department of Energy and research Awards
No.~DE-AC06-76RLO1830, % PNNL
No.~DE-SC0007983, % Wayne State
No.~DE-SC0009824, % Florida
No.~DE-SC0009973, % VPI
No.~DE-SC0010007, % Duke
No.~DE-SC0010073, % South Carolina
No.~DE-SC0010118, % Carnegie Mellon
No.~DE-SC0010504, % Hawaii
No.~DE-SC0011784, % Cincinnati
No.~DE-SC0012704, % BNL
No.~DE-SC0019230, % Duke
No.~DE-SC0021274, % Mississippi
No.~DE-SC0022350; % Louisville
%last group
and
%Vietnam
the Vietnam Academy of Science and Technology (VAST) under Grant No.~DL0000.05/21-23.

% Policy from October 20, 2022
These acknowledgements are not to be interpreted as an endorsement of any statement made
by any of our institutes, funding agencies, governments, or their representatives.

We thank the SuperKEKB team for delivering high-luminosity collisions;
the KEK cryogenics group for the efficient operation of the detector solenoid magnet;
the KEK computer group and the NII for on-site computing support and SINET6 network support;
and the raw-data centers at BNL, DESY, GridKa, IN2P3, INFN, and the University of Victoria for offsite computing support.

\end{acknowledgments}

%%%%%%%%%%%%%%%%%%%%%%%%%
\bibliography{belle2}

\begin{thebibliography}{47}%
\makeatletter
\providecommand \@ifxundefined [1]{%
 \@ifx{#1\undefined}
}%
\providecommand \@ifnum [1]{%
 \ifnum #1\expandafter \@firstoftwo
 \else \expandafter \@secondoftwo
 \fi
}%
\providecommand \@ifx [1]{%
 \ifx #1\expandafter \@firstoftwo
 \else \expandafter \@secondoftwo
 \fi
}%
\providecommand \natexlab [1]{#1}%
\providecommand \enquote  [1]{``#1''}%
\providecommand \bibnamefont  [1]{#1}%
\providecommand \bibfnamefont [1]{#1}%
\providecommand \citenamefont [1]{#1}%
\providecommand \href@noop [0]{\@secondoftwo}%
\providecommand \href [0]{\begingroup \@sanitize@url \@href}%
\providecommand \@href[1]{\@@startlink{#1}\@@href}%
\providecommand \@@href[1]{\endgroup#1\@@endlink}%
\providecommand \@sanitize@url [0]{\catcode `\\12\catcode `\$12\catcode
  `\&12\catcode `\#12\catcode `\^12\catcode `\_12\catcode `\%12\relax}%
\providecommand \@@startlink[1]{}%
\providecommand \@@endlink[0]{}%
\providecommand \url  [0]{\begingroup\@sanitize@url \@url }%
\providecommand \@url [1]{\endgroup\@href {#1}{\urlprefix }}%
\providecommand \urlprefix  [0]{URL }%
\providecommand \Eprint [0]{\href }%
\providecommand \doibase [0]{https://doi.org/}%
\providecommand \selectlanguage [0]{\@gobble}%
\providecommand \bibinfo  [0]{\@secondoftwo}%
\providecommand \bibfield  [0]{\@secondoftwo}%
\providecommand \translation [1]{[#1]}%
\providecommand \BibitemOpen [0]{}%
\providecommand \bibitemStop [0]{}%
\providecommand \bibitemNoStop [0]{.\EOS\space}%
\providecommand \EOS [0]{\spacefactor3000\relax}%
\providecommand \BibitemShut  [1]{\csname bibitem#1\endcsname}%
\let\auto@bib@innerbib\@empty
%</preamble>
\bibitem [{\citenamefont {Lenz}(2015)}]{Lenz:2014jha}%
  \BibitemOpen
  \bibfield  {author} {\bibinfo {author} {\bibfnamefont {A.}~\bibnamefont
  {Lenz}},\ }\href {https://doi.org/10.1142/S0217751X15430058} {\bibfield
  {journal} {\bibinfo  {journal} {Int. J. Mod. Phys. A}\ }\textbf {\bibinfo
  {volume} {30}},\ \bibinfo {pages} {1543005} (\bibinfo {year} {2015})},\
  \Eprint {https://arxiv.org/abs/1405.3601} {arXiv:1405.3601 [hep-ph]}
  \BibitemShut {NoStop}%
\bibitem [{\citenamefont {Dowdall}\ \emph {et~al.}(2019)\citenamefont
  {Dowdall}, \citenamefont {Davies}, \citenamefont {Horgan}, \citenamefont
  {Lepage}, \citenamefont {Monahan}, \citenamefont {Shigemitsu},\ and\
  \citenamefont {Wingate}}]{Dowdall:2019bea}%
  \BibitemOpen
  \bibfield  {author} {\bibinfo {author} {\bibfnamefont {R.~J.}\ \bibnamefont
  {Dowdall}}, \bibinfo {author} {\bibfnamefont {C.~T.~H.}\ \bibnamefont
  {Davies}}, \bibinfo {author} {\bibfnamefont {R.~R.}\ \bibnamefont {Horgan}},
  \bibinfo {author} {\bibfnamefont {G.~P.}\ \bibnamefont {Lepage}}, \bibinfo
  {author} {\bibfnamefont {C.~J.}\ \bibnamefont {Monahan}}, \bibinfo {author}
  {\bibfnamefont {J.}~\bibnamefont {Shigemitsu}},\ and\ \bibinfo {author}
  {\bibfnamefont {M.}~\bibnamefont {Wingate}},\ }\href
  {https://doi.org/10.1103/PhysRevD.100.094508} {\bibfield  {journal} {\bibinfo
   {journal} {Phys. Rev. D}\ }\textbf {\bibinfo {volume} {100}},\ \bibinfo
  {pages} {094508} (\bibinfo {year} {2019})},\ \Eprint
  {https://arxiv.org/abs/1907.01025} {arXiv:1907.01025 [hep-lat]} \BibitemShut
  {NoStop}%
\bibitem [{\citenamefont {Abe}\ \emph {et~al.}(2005)\citenamefont {Abe} \emph
  {et~al.}}]{Belle:2004hwe}%
  \BibitemOpen
  \bibfield  {author} {\bibinfo {author} {\bibfnamefont {K.}~\bibnamefont
  {Abe}} \emph {et~al.} (\bibinfo {collaboration} {Belle Collaboration}),\
  }\href {https://doi.org/10.1103/PhysRevD.71.072003} {\bibfield  {journal}
  {\bibinfo  {journal} {Phys. Rev. D}\ }\textbf {\bibinfo {volume} {71}},\
  \bibinfo {pages} {072003} (\bibinfo {year} {2005})},\ \Eprint
  {https://arxiv.org/abs/hep-ex/0408111} {arXiv:hep-ex/0408111} \BibitemShut
  {NoStop}%
\bibitem [{\citenamefont {Aubert}\ \emph
  {et~al.}(2006{\natexlab{a}})\citenamefont {Aubert} \emph
  {et~al.}}]{BaBar:2005laz}%
  \BibitemOpen
  \bibfield  {author} {\bibinfo {author} {\bibfnamefont {B.}~\bibnamefont
  {Aubert}} \emph {et~al.} (\bibinfo {collaboration} {Babar Collaboration}),\
  }\href {https://doi.org/10.1103/PhysRevD.73.012004} {\bibfield  {journal}
  {\bibinfo  {journal} {Phys. Rev. D}\ }\textbf {\bibinfo {volume} {73}},\
  \bibinfo {pages} {012004} (\bibinfo {year} {2006}{\natexlab{a}})},\ \Eprint
  {https://arxiv.org/abs/hep-ex/0507054} {arXiv:hep-ex/0507054} \BibitemShut
  {NoStop}%
\bibitem [{\citenamefont {Aaij}\ \emph {et~al.}(2014)\citenamefont {Aaij} \emph
  {et~al.}}]{LHCb:2014qsd}%
  \BibitemOpen
  \bibfield  {author} {\bibinfo {author} {\bibfnamefont {R.}~\bibnamefont
  {Aaij}} \emph {et~al.} (\bibinfo {collaboration} {LHCb Collaboration}),\
  }\href {https://doi.org/10.1007/JHEP04(2014)114} {\bibfield  {journal}
  {\bibinfo  {journal} {JHEP}\ }\textbf {\bibinfo {volume} {04}},\ \bibinfo
  {pages} {114}},\ \Eprint {https://arxiv.org/abs/1402.2554} {arXiv:1402.2554
  [hep-ex]} \BibitemShut {NoStop}%
\bibitem [{\citenamefont {Aaij}\ \emph {et~al.}(2016)\citenamefont {Aaij} \emph
  {et~al.}}]{LHCb:2016gsk}%
  \BibitemOpen
  \bibfield  {author} {\bibinfo {author} {\bibfnamefont {R.}~\bibnamefont
  {Aaij}} \emph {et~al.} (\bibinfo {collaboration} {LHCb Collaboration}),\
  }\href {https://doi.org/10.1140/epjc/s10052-016-4250-2} {\bibfield  {journal}
  {\bibinfo  {journal} {Eur. Phys. J. C}\ }\textbf {\bibinfo {volume} {76}},\
  \bibinfo {pages} {412} (\bibinfo {year} {2016})},\ \Eprint
  {https://arxiv.org/abs/1604.03475} {arXiv:1604.03475 [hep-ex]} \BibitemShut
  {NoStop}%
\bibitem [{\citenamefont {Sirunyan}\ \emph {et~al.}(2018)\citenamefont
  {Sirunyan} \emph {et~al.}}]{CMS:2017ygm}%
  \BibitemOpen
  \bibfield  {author} {\bibinfo {author} {\bibfnamefont {A.~M.}\ \bibnamefont
  {Sirunyan}} \emph {et~al.} (\bibinfo {collaboration} {CMS Collaboration}),\
  }\href {https://doi.org/10.1140/epjc/s10052-018-5929-3} {\bibfield  {journal}
  {\bibinfo  {journal} {Eur. Phys. J. C}\ }\textbf {\bibinfo {volume} {78}},\
  \bibinfo {pages} {457} (\bibinfo {year} {2018})},\ \Eprint
  {https://arxiv.org/abs/1710.08949} {arXiv:1710.08949 [hep-ex]} \BibitemShut
  {NoStop}%
\bibitem [{\citenamefont {Aad}\ \emph {et~al.}(2013)\citenamefont {Aad} \emph
  {et~al.}}]{ATLAS:2012cvl}%
  \BibitemOpen
  \bibfield  {author} {\bibinfo {author} {\bibfnamefont {G.}~\bibnamefont
  {Aad}} \emph {et~al.} (\bibinfo {collaboration} {ATLAS Collaboration}),\
  }\href {https://doi.org/10.1103/PhysRevD.87.032002} {\bibfield  {journal}
  {\bibinfo  {journal} {Phys. Rev. D}\ }\textbf {\bibinfo {volume} {87}},\
  \bibinfo {pages} {032002} (\bibinfo {year} {2013})},\ \Eprint
  {https://arxiv.org/abs/1207.2284} {arXiv:1207.2284 [hep-ex]} \BibitemShut
  {NoStop}%
\bibitem [{\citenamefont {Abazov}\ \emph {et~al.}(2015)\citenamefont {Abazov}
  \emph {et~al.}}]{D0:2014ycx}%
  \BibitemOpen
  \bibfield  {author} {\bibinfo {author} {\bibfnamefont {V.~M.}\ \bibnamefont
  {Abazov}} \emph {et~al.} (\bibinfo {collaboration} {D0 Collaboration}),\
  }\href {https://doi.org/10.1103/PhysRevLett.114.062001} {\bibfield  {journal}
  {\bibinfo  {journal} {Phys. Rev. Lett.}\ }\textbf {\bibinfo {volume} {114}},\
  \bibinfo {pages} {062001} (\bibinfo {year} {2015})},\ \Eprint
  {https://arxiv.org/abs/1410.1568} {arXiv:1410.1568 [hep-ex]} \BibitemShut
  {NoStop}%
\bibitem [{\citenamefont {Aaltonen}\ \emph {et~al.}(2011)\citenamefont
  {Aaltonen} \emph {et~al.}}]{CDF:2010ibe}%
  \BibitemOpen
  \bibfield  {author} {\bibinfo {author} {\bibfnamefont {T.}~\bibnamefont
  {Aaltonen}} \emph {et~al.} (\bibinfo {collaboration} {CDF Collaboration}),\
  }\href {https://doi.org/10.1103/PhysRevLett.106.121804} {\bibfield  {journal}
  {\bibinfo  {journal} {Phys. Rev. Lett.}\ }\textbf {\bibinfo {volume} {106}},\
  \bibinfo {pages} {121804} (\bibinfo {year} {2011})},\ \Eprint
  {https://arxiv.org/abs/1012.3138} {arXiv:1012.3138 [hep-ex]} \BibitemShut
  {NoStop}%
\bibitem [{Note1()}]{Note1}%
  \BibitemOpen
  \bibinfo {note} {We use a system of units in which $\protect \hbar =c=1$ and
  mass and frequency have the same dimension.}\BibitemShut {Stop}%
\bibitem [{Note2()}]{Note2}%
  \BibitemOpen
  \bibinfo {note} {Another naming convention, with $\beta \equiv \phi _1$ and
  $\alpha \equiv \phi _2$, is also used in the literature.}\BibitemShut {Stop}%
\bibitem [{\citenamefont {Adachi}\ \emph {et~al.}(2012)\citenamefont {Adachi}
  \emph {et~al.}}]{Belle:2012paq}%
  \BibitemOpen
  \bibfield  {author} {\bibinfo {author} {\bibfnamefont {I.}~\bibnamefont
  {Adachi}} \emph {et~al.} (\bibinfo {collaboration} {Belle Collaboration}),\
  }\href {https://doi.org/10.1103/PhysRevLett.108.171802} {\bibfield  {journal}
  {\bibinfo  {journal} {Phys. Rev. Lett.}\ }\textbf {\bibinfo {volume} {108}},\
  \bibinfo {pages} {171802} (\bibinfo {year} {2012})},\ \Eprint
  {https://arxiv.org/abs/1201.4643} {arXiv:1201.4643 [hep-ex]} \BibitemShut
  {NoStop}%
\bibitem [{\citenamefont {Vanhoefer}\ \emph {et~al.}(2016)\citenamefont
  {Vanhoefer} \emph {et~al.}}]{Belle:2015xfb}%
  \BibitemOpen
  \bibfield  {author} {\bibinfo {author} {\bibfnamefont {P.}~\bibnamefont
  {Vanhoefer}} \emph {et~al.} (\bibinfo {collaboration} {Belle
  Collaboration}),\ }\href {https://doi.org/10.1103/PhysRevD.93.032010}
  {\bibfield  {journal} {\bibinfo  {journal} {Phys. Rev. D}\ }\textbf {\bibinfo
  {volume} {93}},\ \bibinfo {pages} {032010} (\bibinfo {year} {2016})},\
  \bibinfo {note} {[Addendum: Phys.Rev.D 94, 099903 (2016)]},\ \Eprint
  {https://arxiv.org/abs/1510.01245} {arXiv:1510.01245 [hep-ex]} \BibitemShut
  {NoStop}%
\bibitem [{\citenamefont {Beneke}\ \emph {et~al.}(2000)\citenamefont {Beneke},
  \citenamefont {Buchalla}, \citenamefont {Neubert},\ and\ \citenamefont
  {Sachrajda}}]{Beneke:2000ry}%
  \BibitemOpen
  \bibfield  {author} {\bibinfo {author} {\bibfnamefont {M.}~\bibnamefont
  {Beneke}}, \bibinfo {author} {\bibfnamefont {G.}~\bibnamefont {Buchalla}},
  \bibinfo {author} {\bibfnamefont {M.}~\bibnamefont {Neubert}},\ and\ \bibinfo
  {author} {\bibfnamefont {C.~T.}\ \bibnamefont {Sachrajda}},\ }\href
  {https://doi.org/10.1016/S0550-3213(00)00559-9} {\bibfield  {journal}
  {\bibinfo  {journal} {Nucl. Phys. B}\ }\textbf {\bibinfo {volume} {591}},\
  \bibinfo {pages} {313} (\bibinfo {year} {2000})},\ \Eprint
  {https://arxiv.org/abs/hep-ph/0006124} {arXiv:hep-ph/0006124} \BibitemShut
  {NoStop}%
\bibitem [{\citenamefont {Aaij}\ \emph {et~al.}(2018)\citenamefont {Aaij} \emph
  {et~al.}}]{LHCb:2018zap}%
  \BibitemOpen
  \bibfield  {author} {\bibinfo {author} {\bibfnamefont {R.}~\bibnamefont
  {Aaij}} \emph {et~al.} (\bibinfo {collaboration} {LHCb Collaboration}),\
  }\href {https://doi.org/10.1007/JHEP06(2018)084} {\bibfield  {journal}
  {\bibinfo  {journal} {JHEP}\ }\textbf {\bibinfo {volume} {06}},\ \bibinfo
  {pages} {084}},\ \Eprint {https://arxiv.org/abs/1805.03448} {arXiv:1805.03448
  [hep-ex]} \BibitemShut {NoStop}%
\bibitem [{\citenamefont {Aubert}\ \emph
  {et~al.}(2006{\natexlab{b}})\citenamefont {Aubert} \emph
  {et~al.}}]{BaBar:2006slj}%
  \BibitemOpen
  \bibfield  {author} {\bibinfo {author} {\bibfnamefont {B.}~\bibnamefont
  {Aubert}} \emph {et~al.} (\bibinfo {collaboration} {Babar Collaboration}),\
  }\href {https://doi.org/10.1103/PhysRevD.73.111101} {\bibfield  {journal}
  {\bibinfo  {journal} {Phys. Rev. D}\ }\textbf {\bibinfo {volume} {73}},\
  \bibinfo {pages} {111101} (\bibinfo {year} {2006}{\natexlab{b}})},\ \Eprint
  {https://arxiv.org/abs/hep-ex/0602049} {arXiv:hep-ex/0602049} \BibitemShut
  {NoStop}%
\bibitem [{\citenamefont {Bahinipati}\ \emph {et~al.}(2011)\citenamefont
  {Bahinipati} \emph {et~al.}}]{Belle:2011dhx}%
  \BibitemOpen
  \bibfield  {author} {\bibinfo {author} {\bibfnamefont {S.}~\bibnamefont
  {Bahinipati}} \emph {et~al.} (\bibinfo {collaboration} {Belle
  Collaboration}),\ }\href {https://doi.org/10.1103/PhysRevD.84.021101}
  {\bibfield  {journal} {\bibinfo  {journal} {Phys. Rev. D}\ }\textbf {\bibinfo
  {volume} {84}},\ \bibinfo {pages} {021101} (\bibinfo {year} {2011})},\
  \Eprint {https://arxiv.org/abs/1102.0888} {arXiv:1102.0888 [hep-ex]}
  \BibitemShut {NoStop}%
\bibitem [{\citenamefont {Abudin\'en}\ \emph {et~al.}(2022)\citenamefont
  {Abudin\'en} \emph {et~al.}}]{Belle-II:2021zvj}%
  \BibitemOpen
  \bibfield  {author} {\bibinfo {author} {\bibfnamefont {F.}~\bibnamefont
  {Abudin\'en}} \emph {et~al.} (\bibinfo {collaboration} {Belle~II
  Collaboration}),\ }\href {https://doi.org/10.1140/epjc/s10052-022-10180-9}
  {\bibfield  {journal} {\bibinfo  {journal} {Eur. Phys. J. C}\ }\textbf
  {\bibinfo {volume} {82}},\ \bibinfo {pages} {283} (\bibinfo {year} {2022})},\
  \bibinfo {note} {{Here we use the category-based algorithm}},\ \Eprint
  {https://arxiv.org/abs/2110.00790} {arXiv:2110.00790 [hep-ex]} \BibitemShut
  {NoStop}%
\bibitem [{\citenamefont {Bigi}\ and\ \citenamefont
  {Sanda}(1981)}]{Bigi:1981qs}%
  \BibitemOpen
  \bibfield  {author} {\bibinfo {author} {\bibfnamefont {I.~I.~Y.}\
  \bibnamefont {Bigi}}\ and\ \bibinfo {author} {\bibfnamefont {A.~I.}\
  \bibnamefont {Sanda}},\ }\href {https://doi.org/10.1016/0550-3213(81)90519-8}
  {\bibfield  {journal} {\bibinfo  {journal} {Nucl. Phys. B}\ }\textbf
  {\bibinfo {volume} {193}},\ \bibinfo {pages} {85} (\bibinfo {year}
  {1981})}\BibitemShut {NoStop}%
\bibitem [{\citenamefont {Oddone}(1987)}]{Oddone:1987up}%
  \BibitemOpen
  \bibfield  {author} {\bibinfo {author} {\bibfnamefont {P.}~\bibnamefont
  {Oddone}},\ }\bibfield  {booktitle} {\emph {\bibinfo {booktitle} {{UCLA
  Linear-Collider $BB$ Factory Concep. Design: Proceedings}}},\ }\href@noop {}
  {\bibfield  {journal} {\bibinfo  {journal} {eConf}\ }\textbf {\bibinfo
  {volume} {C870126}},\ \bibinfo {pages} {423} (\bibinfo {year}
  {1987})}\BibitemShut {NoStop}%
\bibitem [{\citenamefont {Akai}\ \emph {et~al.}(2018)\citenamefont {Akai},
  \citenamefont {Furukawa},\ and\ \citenamefont {Koiso}}]{Akai:2018mbz}%
  \BibitemOpen
  \bibfield  {author} {\bibinfo {author} {\bibfnamefont {K.}~\bibnamefont
  {Akai}}, \bibinfo {author} {\bibfnamefont {K.}~\bibnamefont {Furukawa}},\
  and\ \bibinfo {author} {\bibfnamefont {H.}~\bibnamefont {Koiso}} (\bibinfo
  {collaboration} {SuperKEKB}),\ }\href
  {https://doi.org/10.1016/j.nima.2018.08.017} {\bibfield  {journal} {\bibinfo
  {journal} {Nucl. Instrum. Meth. A}\ }\textbf {\bibinfo {volume} {907}},\
  \bibinfo {pages} {188} (\bibinfo {year} {2018})},\ \Eprint
  {https://arxiv.org/abs/1809.01958} {arXiv:1809.01958 [physics.acc-ph]}
  \BibitemShut {NoStop}%
\bibitem [{\citenamefont {Abe}\ \emph {et~al.}(2010)\citenamefont {Abe} \emph
  {et~al.}}]{Belle-II:2010dht}%
  \BibitemOpen
  \bibfield  {author} {\bibinfo {author} {\bibfnamefont {T.}~\bibnamefont
  {Abe}} \emph {et~al.} (\bibinfo {collaboration} {Belle~II Collaboration}),\
  }\href@noop {} {\  (\bibinfo {year} {2010})},\ \Eprint
  {https://arxiv.org/abs/1011.0352} {arXiv:1011.0352 [physics.ins-det]}
  \BibitemShut {NoStop}%
\bibitem [{\citenamefont {Kuhr}\ \emph {et~al.}(2019)\citenamefont {Kuhr},
  \citenamefont {Pulvermacher}, \citenamefont {Ritter}, \citenamefont {Hauth},\
  and\ \citenamefont {Braun}}]{Kuhr:2018lps}%
  \BibitemOpen
  \bibfield  {author} {\bibinfo {author} {\bibfnamefont {T.}~\bibnamefont
  {Kuhr}}, \bibinfo {author} {\bibfnamefont {C.}~\bibnamefont {Pulvermacher}},
  \bibinfo {author} {\bibfnamefont {M.}~\bibnamefont {Ritter}}, \bibinfo
  {author} {\bibfnamefont {T.}~\bibnamefont {Hauth}},\ and\ \bibinfo {author}
  {\bibfnamefont {N.}~\bibnamefont {Braun}} (\bibinfo {collaboration} {Belle~II
  Framework Software Group}),\ }\href
  {https://doi.org/10.1007/s41781-018-0017-9} {\bibfield  {journal} {\bibinfo
  {journal} {Comput. Softw. Big Sci.}\ }\textbf {\bibinfo {volume} {3}},\
  \bibinfo {pages} {1} (\bibinfo {year} {2019})},\ \Eprint
  {https://arxiv.org/abs/1809.04299} {arXiv:1809.04299 [physics.comp-ph]}
  \BibitemShut {NoStop}%
\bibitem [{\citenamefont {Bertacchi}\ \emph {et~al.}(2021)\citenamefont
  {Bertacchi} \emph {et~al.}}]{BelleIITrackingGroup:2020hpx}%
  \BibitemOpen
  \bibfield  {author} {\bibinfo {author} {\bibfnamefont {V.}~\bibnamefont
  {Bertacchi}} \emph {et~al.} (\bibinfo {collaboration} {Belle~II Tracking
  Group}),\ }\href {https://doi.org/10.1016/j.cpc.2020.107610} {\bibfield
  {journal} {\bibinfo  {journal} {Comput. Phys. Commun.}\ }\textbf {\bibinfo
  {volume} {259}},\ \bibinfo {pages} {107610} (\bibinfo {year} {2021})},\
  \Eprint {https://arxiv.org/abs/2003.12466} {arXiv:2003.12466
  [physics.ins-det]} \BibitemShut {NoStop}%
\bibitem [{\citenamefont {Jadach}\ \emph {et~al.}(2000)\citenamefont {Jadach},
  \citenamefont {Ward},\ and\ \citenamefont {Was}}]{Jadach:1999vf}%
  \BibitemOpen
  \bibfield  {author} {\bibinfo {author} {\bibfnamefont {S.}~\bibnamefont
  {Jadach}}, \bibinfo {author} {\bibfnamefont {B.~F.~L.}\ \bibnamefont
  {Ward}},\ and\ \bibinfo {author} {\bibfnamefont {Z.}~\bibnamefont {Was}},\
  }\href {https://doi.org/10.1016/S0010-4655(00)00048-5} {\bibfield  {journal}
  {\bibinfo  {journal} {Comput. Phys. Commun.}\ }\textbf {\bibinfo {volume}
  {130}},\ \bibinfo {pages} {260} (\bibinfo {year} {2000})},\ \Eprint
  {https://arxiv.org/abs/hep-ph/9912214} {arXiv:hep-ph/9912214} \BibitemShut
  {NoStop}%
\bibitem [{\citenamefont {Sj\"ostrand}\ \emph {et~al.}(2015)\citenamefont
  {Sj\"ostrand}, \citenamefont {Ask}, \citenamefont {Christiansen},
  \citenamefont {Corke}, \citenamefont {Desai}, \citenamefont {Ilten},
  \citenamefont {Mrenna}, \citenamefont {Prestel}, \citenamefont {Rasmussen},\
  and\ \citenamefont {Skands}}]{Sjostrand:2014zea}%
  \BibitemOpen
  \bibfield  {author} {\bibinfo {author} {\bibfnamefont {T.}~\bibnamefont
  {Sj\"ostrand}}, \bibinfo {author} {\bibfnamefont {S.}~\bibnamefont {Ask}},
  \bibinfo {author} {\bibfnamefont {J.~R.}\ \bibnamefont {Christiansen}},
  \bibinfo {author} {\bibfnamefont {R.}~\bibnamefont {Corke}}, \bibinfo
  {author} {\bibfnamefont {N.}~\bibnamefont {Desai}}, \bibinfo {author}
  {\bibfnamefont {P.}~\bibnamefont {Ilten}}, \bibinfo {author} {\bibfnamefont
  {S.}~\bibnamefont {Mrenna}}, \bibinfo {author} {\bibfnamefont
  {S.}~\bibnamefont {Prestel}}, \bibinfo {author} {\bibfnamefont {C.~O.}\
  \bibnamefont {Rasmussen}},\ and\ \bibinfo {author} {\bibfnamefont {P.~Z.}\
  \bibnamefont {Skands}},\ }\href {https://doi.org/10.1016/j.cpc.2015.01.024}
  {\bibfield  {journal} {\bibinfo  {journal} {Comput. Phys. Commun.}\ }\textbf
  {\bibinfo {volume} {191}},\ \bibinfo {pages} {159} (\bibinfo {year}
  {2015})},\ \Eprint {https://arxiv.org/abs/1410.3012} {arXiv:1410.3012
  [hep-ph]} \BibitemShut {NoStop}%
\bibitem [{\citenamefont {Lange}(2001)}]{Lange:2001uf}%
  \BibitemOpen
  \bibfield  {author} {\bibinfo {author} {\bibfnamefont {D.~J.}\ \bibnamefont
  {Lange}},\ }\href {https://doi.org/10.1016/S0168-9002(01)00089-4} {\bibfield
  {journal} {\bibinfo  {journal} {Nucl. Instrum. Meth. A}\ }\textbf {\bibinfo
  {volume} {462}},\ \bibinfo {pages} {152} (\bibinfo {year}
  {2001})}\BibitemShut {NoStop}%
\bibitem [{\citenamefont {Agostinelli}\ \emph {et~al.}(2003)\citenamefont
  {Agostinelli} \emph {et~al.}}]{GEANT4:2002zbu}%
  \BibitemOpen
  \bibfield  {author} {\bibinfo {author} {\bibfnamefont {S.}~\bibnamefont
  {Agostinelli}} \emph {et~al.} (\bibinfo {collaboration} {GEANT4 software
  Group}),\ }\href {https://doi.org/10.1016/S0168-9002(03)01368-8} {\bibfield
  {journal} {\bibinfo  {journal} {Nucl. Instrum. Meth. A}\ }\textbf {\bibinfo
  {volume} {506}},\ \bibinfo {pages} {250} (\bibinfo {year}
  {2003})}\BibitemShut {NoStop}%
\bibitem [{\citenamefont {Liptak}\ \emph {et~al.}(2022)\citenamefont {Liptak}
  \emph {et~al.}}]{Liptak:2021tog}%
  \BibitemOpen
  \bibfield  {author} {\bibinfo {author} {\bibfnamefont {Z.~J.}\ \bibnamefont
  {Liptak}} \emph {et~al.},\ }\href
  {https://doi.org/10.1016/j.nima.2022.167168} {\bibfield  {journal} {\bibinfo
  {journal} {Nucl. Instrum. Meth. A}\ }\textbf {\bibinfo {volume} {1040}},\
  \bibinfo {pages} {167168} (\bibinfo {year} {2022})},\ \Eprint
  {https://arxiv.org/abs/2112.14537} {arXiv:2112.14537 [physics.ins-det]}
  \BibitemShut {NoStop}%
\bibitem [{\citenamefont {Hulsbergen}(2009)}]{Amoraal:2012qn}%
  \BibitemOpen
  \bibfield  {author} {\bibinfo {author} {\bibfnamefont {W.}~\bibnamefont
  {Hulsbergen}},\ }\href {https://doi.org/10.1016/j.nima.2008.11.094}
  {\bibfield  {journal} {\bibinfo  {journal} {Nucl. Instrum. Meth. A}\ }\textbf
  {\bibinfo {volume} {600}},\ \bibinfo {pages} {471} (\bibinfo {year}
  {2009})},\ \Eprint {https://arxiv.org/abs/0810.2241} {arXiv:0810.2241
  [physics.ins-det]} \BibitemShut {NoStop}%
\bibitem [{\citenamefont {Krohn}\ \emph {et~al.}(2020)\citenamefont {Krohn}
  \emph {et~al.}}]{krohn}%
  \BibitemOpen
  \bibfield  {author} {\bibinfo {author} {\bibfnamefont {J.~F.}\ \bibnamefont
  {Krohn}} \emph {et~al.} (\bibinfo {collaboration} {Belle~II analysis software
  Group}),\ }\href {https://doi.org/10.1016/j.nima.2020.164269} {\bibfield
  {journal} {\bibinfo  {journal} {Nucl. Instrum. Meth. A}\ }\textbf {\bibinfo
  {volume} {976}},\ \bibinfo {pages} {164269} (\bibinfo {year} {2020})},\
  \Eprint {https://arxiv.org/abs/1901.11198} {arXiv:1901.11198 [hep-ex]}
  \BibitemShut {NoStop}%
\bibitem [{\citenamefont {Waltenberger}\ \emph {et~al.}(2008)\citenamefont
  {Waltenberger}, \citenamefont {Mitaroff}, \citenamefont {Moser},
  \citenamefont {Pflugfelder},\ and\ \citenamefont
  {Riedel}}]{Waltenberger:2008zza}%
  \BibitemOpen
  \bibfield  {author} {\bibinfo {author} {\bibfnamefont {W.}~\bibnamefont
  {Waltenberger}}, \bibinfo {author} {\bibfnamefont {W.}~\bibnamefont
  {Mitaroff}}, \bibinfo {author} {\bibfnamefont {F.}~\bibnamefont {Moser}},
  \bibinfo {author} {\bibfnamefont {B.}~\bibnamefont {Pflugfelder}},\ and\
  \bibinfo {author} {\bibfnamefont {H.~V.}\ \bibnamefont {Riedel}},\ }\href
  {https://doi.org/10.1088/1742-6596/119/3/032037} {\bibfield  {journal}
  {\bibinfo  {journal} {J. Phys. Conf. Ser.}\ }\textbf {\bibinfo {volume}
  {119}},\ \bibinfo {pages} {032037} (\bibinfo {year} {2008})}\BibitemShut
  {NoStop}%
\bibitem [{\citenamefont {Dey}\ and\ \citenamefont
  {Soffer}(2020)}]{Dey:2020dsr}%
  \BibitemOpen
  \bibfield  {author} {\bibinfo {author} {\bibfnamefont {S.}~\bibnamefont
  {Dey}}\ and\ \bibinfo {author} {\bibfnamefont {A.}~\bibnamefont {Soffer}},\
  }\href {https://doi.org/10.1007/978-981-15-6292-1_52} {\bibfield  {journal}
  {\bibinfo  {journal} {Springer Proc. Phys.}\ }\textbf {\bibinfo {volume}
  {248}},\ \bibinfo {pages} {411} (\bibinfo {year} {2020})}\BibitemShut
  {NoStop}%
\bibitem [{\citenamefont {Chen}\ and\ \citenamefont
  {Guestrin}(2016)}]{xgboost}%
  \BibitemOpen
  \bibfield  {author} {\bibinfo {author} {\bibfnamefont {T.}~\bibnamefont
  {Chen}}\ and\ \bibinfo {author} {\bibfnamefont {C.}~\bibnamefont
  {Guestrin}},\ }in\ \href {https://doi.org/10.1145/2939672.2939785} {\emph
  {\bibinfo {booktitle} {Proceedings of the 22nd ACM SIGKDD International
  Conference on Knowledge Discovery and Data Mining}}},\ \bibinfo {series and
  number} {KDD '16}\ (\bibinfo  {publisher} {Association for Computing
  Machinery},\ \bibinfo {address} {New York, NY, USA},\ \bibinfo {year}
  {2016})\ p.\ \bibinfo {pages} {785–794}\BibitemShut {NoStop}%
\bibitem [{\citenamefont {Fox}\ and\ \citenamefont {Wolfram}(1978)}]{fw}%
  \BibitemOpen
  \bibfield  {author} {\bibinfo {author} {\bibfnamefont {G.~C.}\ \bibnamefont
  {Fox}}\ and\ \bibinfo {author} {\bibfnamefont {S.}~\bibnamefont {Wolfram}},\
  }\href {https://doi.org/10.1103/PhysRevLett.41.1581} {\bibfield  {journal}
  {\bibinfo  {journal} {Phys. Rev. Lett.}\ }\textbf {\bibinfo {volume} {41}},\
  \bibinfo {pages} {1581} (\bibinfo {year} {1978})}\BibitemShut {NoStop}%
\bibitem [{\citenamefont {Lee}\ \emph {et~al.}(2003)\citenamefont {Lee} \emph
  {et~al.}}]{Belle:ksfw}%
  \BibitemOpen
  \bibfield  {author} {\bibinfo {author} {\bibfnamefont {S.~H.}\ \bibnamefont
  {Lee}} \emph {et~al.} (\bibinfo {collaboration} {Belle Collaboration}),\
  }\href {https://doi.org/10.1103/PhysRevLett.91.261801} {\bibfield  {journal}
  {\bibinfo  {journal} {Phys. Rev. Lett.}\ }\textbf {\bibinfo {volume} {91}},\
  \bibinfo {pages} {261801} (\bibinfo {year} {2003})},\ \Eprint
  {https://arxiv.org/abs/hep-ex/0308040} {arXiv:hep-ex/0308040} \BibitemShut
  {NoStop}%
\bibitem [{\citenamefont {Asner}\ \emph {et~al.}(1996)\citenamefont {Asner}
  \emph {et~al.}}]{CLEO:1995rok}%
  \BibitemOpen
  \bibfield  {author} {\bibinfo {author} {\bibfnamefont {D.~M.}\ \bibnamefont
  {Asner}} \emph {et~al.} (\bibinfo {collaboration} {CLEO Collaboration}),\
  }\href {https://doi.org/10.1103/PhysRevD.53.1039} {\bibfield  {journal}
  {\bibinfo  {journal} {Phys. Rev. D}\ }\textbf {\bibinfo {volume} {53}},\
  \bibinfo {pages} {1039} (\bibinfo {year} {1996})},\ \Eprint
  {https://arxiv.org/abs/hep-ex/9508004} {arXiv:hep-ex/9508004} \BibitemShut
  {NoStop}%
\bibitem [{\citenamefont {{Ed. A.J.~Bevan, B.~Golob, Th.~Mannel, S.~Prell, and
  B.D.~Yabsley}}(2014)}]{BaBar:2014omp}%
  \BibitemOpen
  \bibfield  {author} {\bibinfo {author} {\bibnamefont {{Ed. A.J.~Bevan,
  B.~Golob, Th.~Mannel, S.~Prell, and B.D.~Yabsley}}},\ }\href
  {https://doi.org/10.1140/epjc/s10052-014-3026-9} {\bibfield  {journal}
  {\bibinfo  {journal} {Eur. Phys. J. C}\ }\textbf {\bibinfo {volume} {74}},\
  \bibinfo {pages} {3026} (\bibinfo {year} {2014})},\ \bibinfo {note} {{Chapter
  9}},\ \Eprint {https://arxiv.org/abs/1406.6311} {arXiv:1406.6311 [hep-ex]}
  \BibitemShut {NoStop}%
\bibitem [{\citenamefont {Parisi}(1978)}]{Parisi:1978eg}%
  \BibitemOpen
  \bibfield  {author} {\bibinfo {author} {\bibfnamefont {G.}~\bibnamefont
  {Parisi}},\ }\href {https://doi.org/10.1016/0370-2693(78)90061-8} {\bibfield
  {journal} {\bibinfo  {journal} {Phys. Lett. B}\ }\textbf {\bibinfo {volume}
  {74}},\ \bibinfo {pages} {65} (\bibinfo {year} {1978})}\BibitemShut {NoStop}%
\bibitem [{\citenamefont {Skwarnicki}(1986)}]{Gaiser:1982yw}%
  \BibitemOpen
  \bibfield  {author} {\bibinfo {author} {\bibfnamefont {T.}~\bibnamefont
  {Skwarnicki}},\ }\emph {\bibinfo {title} {{A study of the radiative CASCADE
  transitions between the Upsilon-Prime and Upsilon resonances}}},\ \href@noop
  {} {Ph.D. thesis},\ \bibinfo  {school} {Cracow, INP} (\bibinfo {year}
  {1986})\BibitemShut {NoStop}%
\bibitem [{\citenamefont {Johnson}(1949)}]{johnson}%
  \BibitemOpen
  \bibfield  {author} {\bibinfo {author} {\bibfnamefont {N.~L.}\ \bibnamefont
  {Johnson}},\ }\href@noop {} {\bibfield  {journal} {\bibinfo  {journal}
  {Biometrika}\ }\textbf {\bibinfo {volume} {36}},\ \bibinfo {pages} {149}
  (\bibinfo {year} {1949})}\BibitemShut {NoStop}%
\bibitem [{\citenamefont {Pivk}\ and\ \citenamefont
  {Le~Diberder}(2005)}]{Pivk:2004ty}%
  \BibitemOpen
  \bibfield  {author} {\bibinfo {author} {\bibfnamefont {M.}~\bibnamefont
  {Pivk}}\ and\ \bibinfo {author} {\bibfnamefont {F.~R.}\ \bibnamefont
  {Le~Diberder}},\ }\href {https://doi.org/10.1016/j.nima.2005.08.106}
  {\bibfield  {journal} {\bibinfo  {journal} {Nucl. Instrum. Meth. A}\ }\textbf
  {\bibinfo {volume} {555}},\ \bibinfo {pages} {356} (\bibinfo {year}
  {2005})},\ \Eprint {https://arxiv.org/abs/physics/0402083}
  {arXiv:physics/0402083} \BibitemShut {NoStop}%
\bibitem [{\citenamefont {Dembinski}\ \emph {et~al.}(2021)\citenamefont
  {Dembinski}, \citenamefont {Kenzie}, \citenamefont {Langenbruch},\ and\
  \citenamefont {Schmelling}}]{Dembinski:2021kim}%
  \BibitemOpen
  \bibfield  {author} {\bibinfo {author} {\bibfnamefont {H.}~\bibnamefont
  {Dembinski}}, \bibinfo {author} {\bibfnamefont {M.}~\bibnamefont {Kenzie}},
  \bibinfo {author} {\bibfnamefont {C.}~\bibnamefont {Langenbruch}},\ and\
  \bibinfo {author} {\bibfnamefont {M.}~\bibnamefont {Schmelling}},\
  }\href@noop {} {\  (\bibinfo {year} {2021})},\ \Eprint
  {https://arxiv.org/abs/2112.04574} {arXiv:2112.04574 [stat.ME]} \BibitemShut
  {NoStop}%
\bibitem [{\citenamefont {Efron}(1979)}]{bootstrap}%
  \BibitemOpen
  \bibfield  {author} {\bibinfo {author} {\bibfnamefont {B.}~\bibnamefont
  {Efron}},\ }\href {https://doi.org/10.1214/aos/1176344552} {\bibfield
  {journal} {\bibinfo  {journal} {The Annals of Statistics}\ }\textbf {\bibinfo
  {volume} {7}},\ \bibinfo {pages} {1 } (\bibinfo {year} {1979})}\BibitemShut
  {NoStop}%
\bibitem [{\citenamefont {Bilka}\ \emph {et~al.}(2021)\citenamefont {Bilka},
  \citenamefont {Kandra}, \citenamefont {Kleinwort},\ and\ \citenamefont
  {Zlebcik}}]{Bilka:2021rqj}%
  \BibitemOpen
  \bibfield  {author} {\bibinfo {author} {\bibfnamefont {T.}~\bibnamefont
  {Bilka}}, \bibinfo {author} {\bibfnamefont {J.}~\bibnamefont {Kandra}},
  \bibinfo {author} {\bibfnamefont {C.}~\bibnamefont {Kleinwort}},\ and\
  \bibinfo {author} {\bibfnamefont {R.}~\bibnamefont {Zlebcik}},\ }\href
  {https://doi.org/10.1051/epjconf/202125103028} {\bibfield  {journal}
  {\bibinfo  {journal} {EPJ Web Conf.}\ }\textbf {\bibinfo {volume} {251}},\
  \bibinfo {pages} {03028} (\bibinfo {year} {2021})}\BibitemShut {NoStop}%
\bibitem [{\citenamefont {Workman}\ \emph {et~al.}(2022)\citenamefont {Workman}
  \emph {et~al.}}]{PDG}%
  \BibitemOpen
  \bibfield  {author} {\bibinfo {author} {\bibfnamefont {R.~L.}\ \bibnamefont
  {Workman}} \emph {et~al.} (\bibinfo {collaboration} {Particle Data Group}),\
  }\href {https://doi.org/10.1093/ptep/ptac097} {\bibfield  {journal} {\bibinfo
   {journal} {PTEP}\ }\textbf {\bibinfo {volume} {2022}},\ \bibinfo {pages}
  {083C01} (\bibinfo {year} {2022})}\BibitemShut {NoStop}%
\end{thebibliography}%
\bibliographystyle{apsrev4-2.bst}

\end{document}